\documentclass[11pt]{article}
\usepackage{times}
\usepackage{amssymb}
\usepackage{amsfonts}

\newtheorem{theorem}{Theorem}
\newcommand{\bt}{\begin{theorem}}
\newcommand{\et}{\end{theorem}}
\newtheorem{definition}{Definition}
\newcommand{\bd}{\begin{definition}}
\newcommand{\ed}{\end{definition}}
\newtheorem{proposition}{Proposition}
\newcommand{\bp}{\begin{proposition}}
\newcommand{\ep}{\end{proposition}}
\newtheorem{corollary}{Corollary}
\newcommand{\bc}{\begin{corollary}}
\newcommand{\ec}{\end{corollary}}
\newtheorem{example}{Example}
\newcommand{\be}{\begin{example}}
\newcommand{\ee}{\end{example}}
\newtheorem{lemma}{Lemma}
\newcommand{\bl}{\begin{lemma}}
\newcommand{\el}{\end{lemma}}
\newtheorem{remark}{Remark}
\newcommand{\rk}{\begin{remark}}
\newcommand{\mk}{\end{remark}}
\newcommand{\beq}[1]{\begin{equation} \label{#1}}

\renewcommand{\textwidth}{6.5 in}

\renewcommand{\baselinestretch}{1.1}

\usepackage{latexsym}
  \usepackage{rotate}
\usepackage{amsmath,amssymb}
\usepackage{epsf,epsfig}

\newcommand{\qed}{ \hfill $\square$ \vspace{3mm}}

\allowdisplaybreaks[1]

\newcommand{\ddoublespace}{\renewcommand{\baselinestretch}{1.5}\small\normalsize}
\newcommand{\Doublespace}{\renewcommand{\baselinestretch}{1.1}\small\normalsize}

\begin{document}

\ddoublespace

\begin{center}
\begin{tabular}[b]{c}
{\LARGE \bf Mathematics and Engineering} \\ \medskip
{\LARGE \bf Communications Laboratory}
\ \\
\ \\
{\Large \sf Technical Report}
\end{tabular}\hspace{0.4in}\epsfig{file=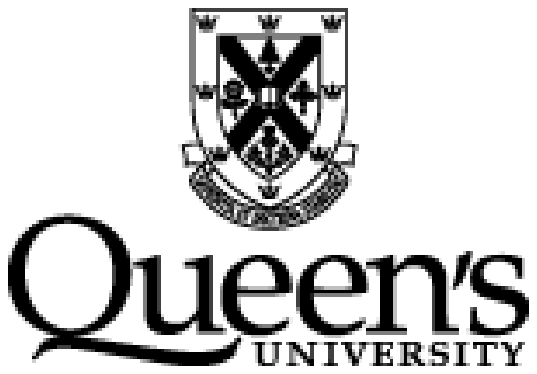,width=3.0cm}

\rule{6.0in}{0.8mm}
\end{center}

\bigskip
\bigskip
\bigskip
\bigskip
\vspace{1.0in}

\begin{center}
{\Large\bf Inner and Outer Bounds for the Public Information} \\
\smallskip
{\Large\bf Embedding Capacity Region Under Multiple Access Attacks} \\
\vspace{0.6in}
{\large {\it Y. Zhong}, {\it Y. Wang}, {\it F. Alajaji}, and {\it T. Linder}}  \\
\end{center}

\vspace{2.0in}
\begin{center}
{\large March 2010}
\end{center}

\newpage

\Doublespace

\title {\bf Inner and Outer Bounds for the Public Information Embedding Capacity Region
Under Multiple Access Attacks\footnote {This research was
supported in part by the Natural Sciences and Engineering Research
Council of Canada (NSERC). The material in this correspondence was
presented in part at the IEEE International Symposium on
Information Theory, Toronto, July 2008. \protect
\\ \indent Yangfan Zhong and Yadong Wang are with the Bank of Montreal, 8th
floor, 302 Bay St., Toronto, Canada (email: zhongyangfan@hotmail.com, y.d.wang99@gmail.com);
Fady Alajaji and Tam\'{a}s Linder are with the
Department of Mathematics \& Statistics, Queen's University,
Kingston, ON K7L 3N6, Canada (email:
\{fady,linder\}@mast.queensu.ca). 
}}

\bigskip

\author{Yangfan Zhong  \and Yadong Wang \and Fady Alajaji  \and Tam\'{a}s Linder}

\bigskip

\maketitle

\begin{abstract}

We consider a public multi-user information embedding (watermarking)
system in which two messages (watermarks) are independently embedded
into two correlated covertexts and are transmitted through a
multiple-access attack channel. The tradeoff between the achievable
embedding rates and the average distortions for the two embedders is
studied. For given distortion levels, inner and outer bounds for the
embedding capacity region are obtained in single-letter form.
Tighter bounds are also given for independent covertexts.

\end{abstract}

\vspace{5mm}
\bigskip



\section{Introduction}

In the last decade, the single-user (point-to-point)
information-hiding (information-embedding, watermarking) model has
been thoroughly studied from an information-theoretic point of view;
see, e.g.,  \cite{Cohen02,Moulin03,Willems} and the references therein. With
the rapid development of wired and wireless communication networks,
situations arise where privacy protection is no longer a
point-to-point problem. Therefore, it is of interest to study
information-hiding problems in multi-user settings.

In this paper  we consider the scenario in which two secret messages
(watermarks) are independently embedded in two correlated sources
(covertexts) and are then jointly decoded under multiple-access
attacks. This scenario is motivated by, for example, the practical
situation where audio and video frames are watermarked separately,
but they are transmitted in a single bit stream and decoded by one
multimedia player (see \cite{Moulin04,Sun04,Kotagiri07}). The model is depicted in
Fig.~\ref{model} and it assumes that  two users separately embed their
watermarks 
$W_1$ and $W_2$ into two correlated discrete memoryless sources
(DMSs), $U_1$ and $U_2$. Each user can only access one of the two
covertexts. The watermarked messages (stegotexts) $X^n_1$ and
$X^n_2$ are then sent through a multiple-access attack channel
(MAAC) to a decoder which attempts to reconstruct the watermarks.
For this two-user information embedding system  we are interested in
determining the embedding capacity region; i.e., the two-dimensional
set of all achievable embedding rate pairs under constraints on the
embedding distortions.

\bigskip

\begin{center}

\begin{figure}[h!]
\centering \setlength{\unitlength}{0.09in}
\begin{picture}(55,28)
\put(1,7.5){\footnotesize$U^n_2$} \put(1,23.5){\footnotesize$U^n_1$}
\put(1,7){\vector(1,0){6}} \put(1,23){\vector(1,0){6}}

\put(7,21){\framebox(12,4)} \put(9,22.8){\footnotesize{Encoder $f_1^{(n)}$}}
\put(7,5){\framebox(12,4)} \put(9,6.8){\footnotesize{Encoder $f_2^{(n)}$}} 
\put(13,2){\vector(0,1){3}} \put(13,28){\vector(0,-1){3}}
\put(12.5,0.5){\footnotesize$W_2$}
\put(12.5,28.5){\footnotesize$W_1$}

\put(19,23){\vector(1,0){6}} \put(25,21){\framebox(10,4)}
\put(26.5,22.8){\footnotesize{Destination}}
\put(19,7){\vector(1,0){6}}
\put(25,5){\framebox(10,4)}\put(26.5,6.8){\footnotesize{Destination}}

\put(23,7){\line(0,1){6}} \put(23,23){\line(0,-1){6}}
\put(23,13){\vector(1,0){2}} \put(23,17){\vector(1,0){2}}

\put(20,23.5){\footnotesize$X^n_1$} 
\put(20,7.5){\footnotesize$X^n_2$}

\put(25,11){\framebox(10,8)} \put(28,16){\footnotesize{MAAC}}
\put(28,14){\footnotesize{$W_{Y|X_1X_2}$}}

\put(35,15){\vector(1,0){5}} \put(36,15.5){\footnotesize$Y^n$} 
\put(40,11){\framebox(8,8)} \put(42,16.5){\footnotesize{Joint}} \put(41,14.8){\footnotesize{Decoder}} \put(42.5,13){\footnotesize{$\psi^{(n)}$}}

\put(48,15){\vector(1,0){8}}
\put(49,15.6){\footnotesize$(\widehat{W}_1,\widehat{W}_2)$}

\end{picture}

\vspace{-0.2cm}\caption{A multi-user information embedding system
with two embedders.} \label{model}
\end{figure}
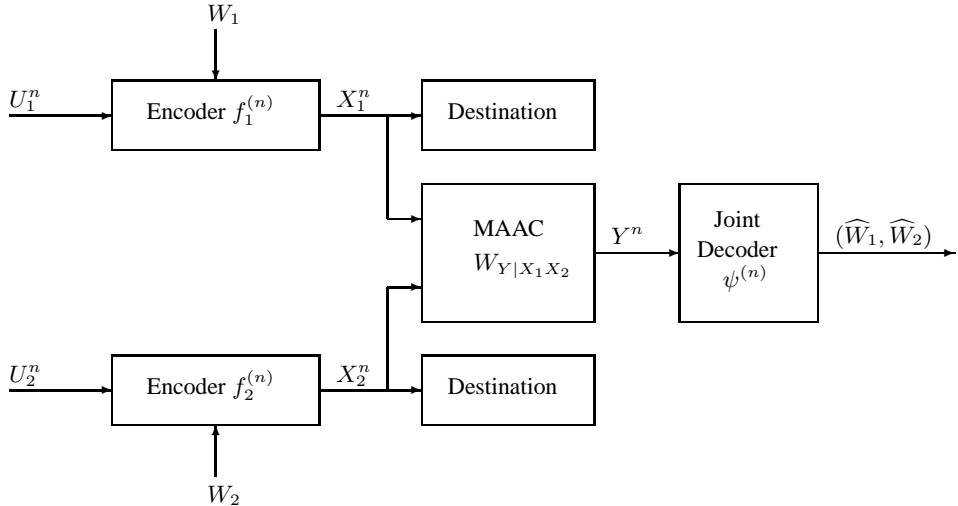

\end{center}

\vspace{-0.3cm}

Our main result (Theorem \ref{theo1}) is an inner bound for the
embedding capacity region. The proof is based on the approach of
Gelfand and Pinsker \cite{GelPin} and a strong typicality
coding/decoding argument. The encoders first map the watermarks $W_1$
and $W_2$ and the correlated covertexts $U^n_1$ and $U^n_2$ to
auxiliary codewords $T^n_1$ and $T^n_2$, and then generate two
stegotexts $X^n_1$ and $X^n_2$ which are jointly typical with
$(U^n_1,U^n_2,T^n_1,T^n_2)$.  The decoder recovers the watermarks by
examining the joint typicality of the received sequence $Y^n$ and all
auxiliary codeword pairs $(T^n_1,T^n_2)$.

One major technical difficulty is the problem of how to separately
construct the typical sequence encoders. In order to guarantee that
the codewords together with the covertexts are jointly typical with
a high probability, we adopt a ``Markov'' encoding scheme from
\cite{Oohama97}, which was originally proposed for Gaussian
multi-terminal source coding (see also \cite{Tung_thesis} and
\cite{HanKob80}). The Markov encoders can be briefly described as
follows. One of the encoders (embedders), say Encoder 1, first forms
an estimate of the source sequence of the other encoder, and then
generates $T^n_1$ which is jointly typical with the observed source
sequence $U^n_1$ and the estimated source sequence. The other
encoder, Encoder 2, first forms an estimate of the source sequence
as well as the auxiliary codeword of Encoder 1, and then generates
$T^n_2$ which is jointly typical with the source sequence $U^n_2$
and all the other sequences estimated. For the resulting scheme, an
extended Markov lemma (Lemma \ref{lemma_EMK}) ensures that the
auxiliary codewords $T^n_1$ and $T^n_2$, although generated by
separate encoders, are jointly typical with the source sequences
with a high probability.

We also derive an outer bound for the embedding capacity region with
single-letter characterization (Theorem \ref{theo2}), using Fano's
inequality and a standard information-theoretical bounding argument.
We specialize  the embedding capacity region to independent covertexts
and obtain inner and outer bounds for
this case (Theorem \ref{theo3}). The inner bound is a consequence of
Theorem \ref{theo1}, while in the converse part we sharpen the bound
of Theorem \ref{theo2} by making use of the independence condition.

We note that the multi-user information embedding problem
studied in this paper is related to the works \cite{Sun04} and
\cite{Kotagiri07}. In \cite{Sun04}, the authors present an
achievable embedding region for correlated Gaussian covertexts and
parallel (independent) additive Gaussian attack channels (as opposed
to the MAAC considered here). In a recent work \cite{Kotagiri07},
the authors study the same system as ours and give an inner bound
for the capacity region without a proof, stating that this inner
bound can be easily proved via the coding procedure in \cite{Sun04}.
However, the proof in \cite{Sun04} seems to be incorrect because the
encoders cannot guarantee the typicality of the output sequences
with respect to the covertexts sequences. Our code construction
corrects this problem and in Theorem \ref{theo1} we show that the
main result in \cite{Sun04} (the achievable region) and the inner
bound given in \cite{Kotagiri07} are both correct.  We also point out
that a similar setup concerning a 
multi-user reversible information embedding system was considered in
\cite{Kotagiri} and \cite{Kotagiri07} for two covertexts and a MAAC.
Since in the reversible information embedding problem the secret
messages and the covertexts are both reconstructed at the decoder,
Gelfand and Pinsker coding is not required and the coding strategy
is fundamentally different from ours.

The remainder of this paper is organized as follows. We set up the
public multi-user embedding (watermarking) problem, define the
embedding capacity region, and present our main results in Section
\ref{mainresults}. The proof of the inner bound is given in
Section~\ref{proof1}, while the proof of the outer bounds are deferred
to the Appendix. We close the paper with concluding remarks in Section
\ref{secCon}.

\section{Problem Formulation and Main Results}\label{mainresults}

Let $|\mathcal{X}|$ denote the size of a finite set $\mathcal{X}$.
If $X$ is a random variable (RV) with distribution $P_{X}$, we
denote its $n$-dimensional product distribution by $P_{X}^{(n)}$.
Similar notation applies to joint and conditional distributions. For
RVs $X$, $Y$, and $Z$ with joint distribution $P_{XYZ}$, we use
$P_X$, $P_{XY}$, $P_{YZ|X}$, etc., to denote the corresponding
marginal and conditional probabilities induced by $P_{XYZ}$. The
expectation of the RV $X$ is denoted by $\mathbb{E}(X)$. All
alphabets are finite, and all logarithms and exponentials are in
base 2.

Let $U_1$ and $U_2$ be two discrete memoryless host sources with
alphabets $\mathcal{U}_1$ and $\mathcal{U}_2$ and joint distribution
$Q_{U_1U_2}$. The watermarks $W_1$ and $W_2$ are independently and
uniformly chosen from the sets
$\mathcal{W}_1\triangleq\{1,2,...,M_1\}$ and
$\mathcal{W}_2\triangleq\{1,2,...,M_2\}$, respectively. The attack
channel is modeled as a two-sender one-receiver discrete memoryless
MAAC $W_{Y|X_1X_2}$ having input alphabets $\mathcal{X}_1$ and
$\mathcal{X}_2$, output alphabet $\mathcal{Y}$, and transition
probability distribution $W_{Y|X_1X_2}(y|x_1,x_2)$. The probability
of receiving $\textbf{y}\in\mathcal{Y}^n$ conditioned on sending
$\textbf{x}_1\in\mathcal{X}_1^n$ and
$\textbf{x}_2\in\mathcal{X}_2^n$ is hence given by
$W_{Y|X_1X_2}^{(n)}(\textbf{y}|\textbf{x}_1,\textbf{x}_2)$.

Let $d_i: \mathcal{U}_i \times \mathcal{X}_i \rightarrow [0,\infty)$
be single-letter distortion measures and define $d_i^{max}
\triangleq \max\limits_{u_i,x_i}d_i(u_i,x_i)$ for $i=1,2$. For
$\textbf{u}_i\in \mathcal{U}_i^n$ and $\textbf{x}_i \in
\mathcal{X}_i^n$, let $d_i(\textbf{u}_i,\textbf{x}_i)= \sum_{j=1}^n
d_i(u_{ij},x_{ij})$.

A two-sender one-receiver multiple-access embedding (MAE) code
$(f_1^{(n)}, f_2^{(n)}, \psi^{(n)})$ with block length $n$ consists
of (see Fig.\ \ref{model}) two encoders (embedders)
$$
f_1^{(n)}:
\mathcal{W}_1\times\mathcal{U}_1^n\longrightarrow\mathcal{X}_1^n\quad
\mbox{ and }\quad
f_2^{(n)}:\mathcal{W}_2\times\mathcal{U}_2^n\longrightarrow\mathcal{X}_2^n
$$
with embedding rates $R_{f_1}=\frac{1}{n}\log_2 M_1$ and
$R_{f_2}=\frac{1}{n}\log_2 M_2$, respectively, and a decoder
$$
\psi^{(n)}:\mathcal{Y}^n\longrightarrow\mathcal{W}_1\times\mathcal{W}_2.
$$

The system depicts a ``public'' embedding scenario since the
covertexts are not available at the decoder. The probability of
erroneously decoding the secret messages is given by
\begin{eqnarray}
P_e^{(n)}&\triangleq&  \Pr\bigl(\psi^{(n)}(Y^n)\neq (W_1,W_2)\bigr) \nonumber\\
&= &\frac{1}{2^{n(R_1+R_2)}}\sum_{w_1=1}^{M_1}
\sum_{w_2=1}^{M_2}\sum_{\mathcal{U}_1^n\times\mathcal{U}_2^n}    
Q_{U_1U_2}^{(n)}(\textbf{u}_1,\textbf{u}_2)
W_{Y|X_1X_2}^{(n)}\left(\textbf{y}:\psi^{(n)}(\textbf{y})\neq
(w_1,w_2)|\textbf{x}_1,\textbf{x}_2\right)\nonumber
\end{eqnarray}
where $\textbf{x}_i\triangleq f_i^{(n)}(w_i,\textbf{u}_i)$ for
$i=1,2$.

\begin{definition}
{\rm Given $Q_{U_1U_2}$, $W_{Y|X_1X_2}$, a rate pair $(R_1,R_2)$ is
said to be achievable with respect to distortion levels $(D_1,D_2)$
if there exists a sequence of MAE codes $(f_1^{(n)},f_2^{(n)},
\psi^{(n)})$ at embedding rates no smaller than $R_1$ and $R_2$,
respectively, such that $\lim_{n\to \infty} P_e^{(n)}= 0$ and
$$
\limsup_{n\to \infty}
\frac{1}{n}\mathbb{E}\left[d_i(U^n_i,f_i^{(n)}(W_i,U^n_i))\right]
\leq D_i, \quad i=1,2.
$$
The embedding capacity region $\mathcal{R}(D_1,D_2)$ is the
closure of the set of all achievable rate pairs $(R_1,R_2)$. }
\end{definition}

\begin{remark}
{\rm It can be shown by using a time-sharing argument
\cite{Cover_book} that $\mathcal{R}(D_1,D_2)$ is convex.}
\end{remark}

\begin{definition} \label{def2}
{\rm Given $Q_{U_1U_2}$, $W_{Y|X_1X_2}$, and a pair of distortion
levels $(D_1,D_2)$, let $\mathcal{S}_{D_1,D_2}$ be the set of RVs
$(U_1,T_1,U_2,T_2,X_1,X_2,Y)\in
\mathcal{U}_1\times\mathcal{T}_1\times\mathcal{U}_2\times\mathcal{T}_2\times\mathcal{X}_1\times\mathcal{X}_2\times\mathcal{Y}$
for some finite alphabets $\mathcal{T}_1$ and $\mathcal{T}_2$ such
that the joint distribution $P_{U_1T_1U_2T_2X_1X_2Y}$ satisfies: (1)
$P_{U_1T_1U_2T_2X_1X_2Y}=Q_{U_1U_2}P_{T_1X_1|U_1}P_{T_2X_2|U_2}W_{Y|X_1X_2},
$ (2) $I(U_i;T_i)>0$, and (3) $\mathbb{E}[d_i(U_i,X_i)] \leq D_i$,
for $i=1,2.$ }
\end{definition}


\begin{definition}
{\rm Given $Q_{U_1U_2}$, $W_{Y|X_1X_2}$, and a pair of distortion
levels $(D_1,D_2)$, let $\mathcal{P}_{D_1,D_2}$ be the set of RVs
$(U_1,T_1,U_2,T_2,X_1,X_2,Y)\in
\mathcal{U}_1\times\mathcal{T}_1\times\mathcal{U}_2\times\mathcal{T}_2\times\mathcal{X}_1\times\mathcal{X}_2\times\mathcal{Y}$
for some finite alphabets $\mathcal{T}_1$ and $\mathcal{T}_2$ such
that the joint distribution $P_{U_1T_1U_2T_2X_1X_2Y}$ satisfies: (1)
$P_{U_1T_1U_2T_2X_1X_2Y}=Q_{U_1U_2}P_{T_1T_2X_1X_2|U_1U_2}W_{Y|X_1X_2},
$ and (2) $\mathbb{E}[d_i(U_i,X_i)] \leq D_i$, for $i=1,2.$ }
\end{definition}

Note that the only difference 
between the two regions is that in the definition of
$\mathcal{S}_{D_1,D_2}$, the conditional distribution of
$(T_1,T_2,X_1,X_2)$ given $(U_1,U_2)$ is restricted to be in the form
$P_{T_1X_1|U_1}P_{T_2X_2|U_2}$. This of course implies
$\mathcal{S}_{D_1,D_2}\subseteq \mathcal{P}_{D_1,D_2}$. 

\smallskip

The following are the main results of the
paper.

\begin{theorem}[Inner bound] \label{theo1}
{\rm Let $\mathcal{R}_{in}(D_1,D_2)$ be the closure of the convex
hull of all $(R_1,R_2)$ satisfying
\begin{eqnarray}
R_1&<& I(T_1;T_2,Y)-I(U_1;T_1),\label{TH1_F1}\\
R_2&<& I(T_2;T_1,Y)-I(U_2;T_2),\label{TH1_F2}\\
R_1+R_2&<& I(T_1,T_2;Y)-I(U_1,U_2;T_1,T_2),\label{TH1_F3}
\end{eqnarray}
for some $(U_1,T_1,U_2,T_2,X_1,X_2,Y)\in \mathcal{S}_{D_1,D_2}$.
Then $\mathcal{R}_{in}(D_1,D_2)\subseteq\mathcal{R}(D_1,D_2)$. }
\end{theorem}

The proof of the theorem is given in  Section \ref{proof1}.

\begin{remark}\label{carsize1}
{\rm As we show in Appendix~\ref{boundcar}, the cardinality of the
  alphabets of the auxiliary RVs $T_1$ and $T_2$ for
  $\mathcal{R}_{in}(D_1,D_2)$ can be bounded as $|\mathcal{T}_i|\leq
  |\mathcal{U}_1||\mathcal{U}_2||\mathcal{X}_i|+1$, $i=1,2$.}
\end{remark}

\begin{remark}\label{remark3}
{\rm Although we only deal with discrete (finite-alphabet) sources
and channels, it is not hard to see that, with the appropriate
changes in the proof, the achievable region is also valid for a
system that incorporates a pair of correlated memoryless Gaussian
sources and a Gaussian MAAC. In particular, when the MAAC is a pair
of parallel (independent) additive Gaussian channels,
$\mathcal{R}_{in}(D_1,D_2)$ is the achievable region
obtained in \cite{Sun04}, even though the proof provided in
\cite{Sun04} is not entirely correct. Note also that our inner bound
$\mathcal{R}_{in}(D_1,D_2)$ is the same as the one given without
proof in \cite[Proposition~1]{Kotagiri07}.}
\end{remark}

\begin{theorem}[Outer bound] \label{theo2}
{\rm Let $\mathcal{R}_{out}(D_1,D_2)$ be the closure of the collection
  of all rate pairs  $(R_1,R_2)$ satisfying conditions
  (\ref{TH1_F1})--(\ref{TH1_F3}) 
for some $(U_1,T_1,U_2,T_2,X_1,X_2,Y)\in \mathcal{P}_{D_1,D_2}$.
Then
$\mathcal{R}(D_1,D_2)\subseteq\mathcal{R}_{out}(D_1+\delta,D_2+\delta)$
for all $\delta>0$. }
\end{theorem}

The proof of the theorem  is given in Appendix
\ref{proof2}. The proof involves Fano's inequality and a
(by now) rather standard information-theoretic argument that
generalizes the converse  proof for a single-user embedding system in
\cite{Willems}. 

\begin{remark}
{\rm The above theorem states that $\mathcal{R}(D_1,D_2)\subseteq
  \bigcap_{\delta>0} \mathcal{R}_{out}(D_1+\delta,D_2+\delta)$. If we
  could upper bound the cardinality of the alphabet sizes of the
  auxiliary RVs $T_1$ and $T_2$ in the definition of
  $\mathcal{R}_{out}(D_1,D_2)$, it would be easy to show that
  $\bigcap_{\delta>0} \mathcal{R}_{out}(D_1+\delta,D_2+\delta) =
  \mathcal{R}_{out}(D_1,D_2)$, so that
  $\mathcal{R}(D_1,D_2)\subseteq\mathcal{R}_{out}(D_1,D_2)$. However,
  without  such an upper bound, we can only state the theorem in the
  present weaker form. The same remark applies to the outer bound in
  the next theorem.}
\end{remark}


We next consider the special case when the covertexts are
independent; i.e., $Q_{U_1U_2}=Q_{U_1}Q_{U_2}$. We then have the
following inner and outer bounds.

\begin{theorem}\label{theo3}
{\rm Let $Q_{U_1U_2}=Q_{U_1}Q_{U_2}$. Let
$\mathcal{R}^*_{in}(D_1,D_2)$ be the closure of the convex hull of
all $(R_1,R_2)$ satisfying
\begin{eqnarray}
R_1&<& I(T_1;Y|T_2)-I(U_1;T_1)\label{TH3_1}\\
R_2&<& I(T_2;Y|T_1)-I(U_2;T_2)\label{TH3_2}\\
R_1+R_2&<& I(T_1,T_2;Y)-I(U_1;T_1)-I(U_2;T_2)\label{TH3_3}
\end{eqnarray}
for some $(U_1,T_1,U_2,T_2,X_1,X_2,Y)\in \mathcal{S}_{D_1,D_2}$, and
let $\mathcal{R}^*_{out}(D_1,D_2)$ be the closure of all $(R_1,R_2)$
satisfying (\ref{TH3_1})--(\ref{TH3_3}) for some
$(U_1,T_1,U_2,T_2,X_1,X_2,Y)\in \mathcal{P}_{D_1,D_2}$. Then
$$
\mathcal{R}^*_{in}(D_1,D_2)\subseteq\mathcal{R}(D_1,D_2)\subseteq\mathcal{R}^*_{out}(D_1+\delta,D_2+\delta)
$$
for all $\delta>0$.
 }
\end{theorem}

The proof is given in Appendix \ref{proof3}. 

\begin{remark}\label{carsize2}
{\rm The cardinality of the alphabets of the auxiliary RVs $T_1$ and
$T_2$ for $\mathcal{R}^*_{in}(D_1,D_2)$ can be bounded as
$|\mathcal{T}_i|\leq |\mathcal{U}_i||\mathcal{X}_i|+1$, $i=1,2$; see
Appendix \ref{boundcar}.}
\end{remark}

\begin{remark}
{\rm In the simple case of independent covertexts
$Q_{U_1U_2}=Q_{U_1}Q_{U_2}$ and parallel MAAC
$W_{Y|X_1X_2}=W_{Y_1|X_1}W_{Y_2|X_2}$ (where
$\mathcal{Y}=\mathcal{Y}_2\times\mathcal{Y}_2$), the inner and outer
bounds of Theorem \ref{theo3} coincide and reduce to the capacity
formula of two parallel single-user watermarking systems
\cite{Moulin03}, \cite{Willems}. }
\end{remark}

\begin{center}

\begin{figure}[h!] \label{fig2}
\centering
\includegraphics[width=10cm]{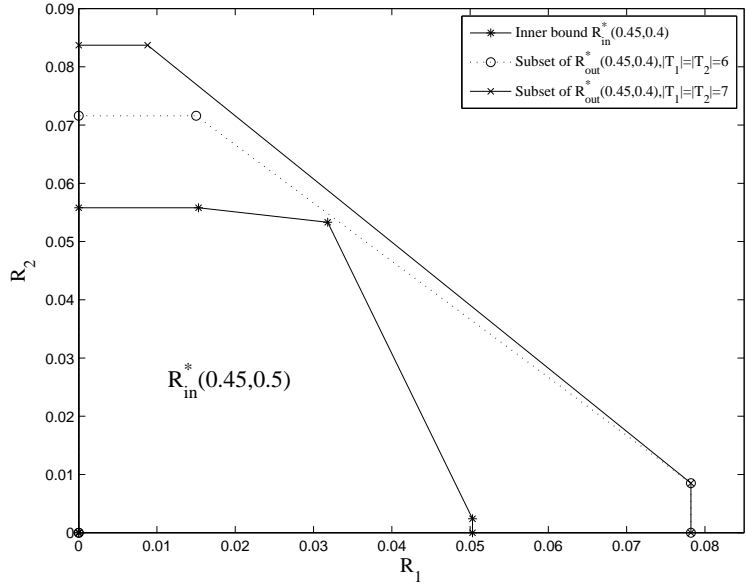}
\caption{The inner bound $\mathcal{R}^*_{in}(0.45,0.4)$ for the
  Example and two subsets of $\mathcal{R}^*_{out}(0.45,0.4)$ obtained
  by setting $|\mathcal{T}_1|=|\mathcal{T}_2|=6$ and
  $|\mathcal{T}_1|=|\mathcal{T}_2|=7$. The obtained regions lie
  between  the corresponding   solid  or dashed  lines and the
  horizontal and vertical 
  axes.  } \label{regionD045D04}
\end{figure}

\end{center}

\noindent{\bf Example}\ \ Let the covertexts be independent binary
sources with $\mathcal{U}_1=\mathcal{U}_2=\{0,1\}$ and
$Q_{U_1}(U_1=0)=0.05$ and $Q_{U_2}(U_2=0)=0.1$. Let the MAAC be a
binary additive channel with
$\mathcal{X}_1=\mathcal{X}_2=\mathcal{Y}=\mathcal{Z}=\{0,1\}$ and
$Y=X_1\oplus X_2\oplus Z$, where $Z$ is independent of $(X_1,X_2)$
with $\Pr(Z=1)=0.02$ and $\oplus$ denotes modulo 2 addition. Let
$D_1=0.45$ and $D_2=0.4$. Fig.~\ref{regionD045D04} illustrates the
numerically computed inner and outer regions of Theorems~\ref{theo1}
and \ref{theo2} (which coincide with the regions of
Theorem~\ref{theo3} since $U_1$ and $U_2$ are independent).  To
compute $\mathcal{R}^*_{in}(0.45,0.4)$, we only need to consider
auxiliary RVs with alphabets $|\mathcal{T}_1|=|\mathcal{T}_2|=5$. For
comparison, we also plot two subsets of the region
$\mathcal{R}^*_{out}(0.45,0.4)$ by setting
$|\mathcal{T}_1|=|\mathcal{T}_2|=6$ and
$|\mathcal{T}_1|=|\mathcal{T}_2|=7$, respectively (recall that
Theorem~\ref{theo3} does not give an upper bound on the alphabet sizes
for $T_2$ and $T_2$ for the outer bound). It is seen that there exist
noticeable gaps between $\mathcal{R}^*_{in}(0.45,0.4)$ and the
numerically obtained subsets of $\mathcal{R}^*_{out}(0.45,0.4)$.
When computing the above regions, we quantized the unit interval using
a step-size of resolution 0.1 to calculate the joint distributions.
We can conclude that the obtained inner and outer bounds do not coincide,
and furthermore, that in case there exists a finite upper bound on the
auxiliary RV alphabet sizes for the outer region, this upper bound
must be at least $7$ for the binary problem.

\section{Proof of Theorem \ref{theo1}} \label{proof1}

We first recall some notation and facts regarding strongly
$\epsilon$-typicality.
Let $V\triangleq(X_1,X_2,...,X_m)$ be a superletter (a collection of
RVs) taking values in a finite set
$\mathcal{V}\triangleq\mathcal{X}_1\times\mathcal{X}_2\times\cdots\times\mathcal{X}_m$
and having joint distribution $P_{V}(x_1,...,x_m)$, which for
simplicity we also denote by $P_V(v)$. Denote by
$T_{\epsilon}^{(n)}(V)$ or $T_{\epsilon}^{(n)}$ the set of all
strongly $\epsilon$-typical sequences \cite[p.~326]{Cover_book} with
respect to the joint distribution $P_V(v)$. Let
$I_V\triangleq\{1,2,...,m\}$, and $I_{G}\subseteq I_V$. We then let
$G=(X_{g_1},X_{g_2},...,X_{g_{|I_G|}}) \in \mathcal{G}$ be a
``sub-superletter'' corresponding to $I_G$ such that $g_i\in I_{G}$.
Let $G$, $K$, and $L$ be sub-superletters of $V$ such that $I_G$,
$I_K$, $I_L$ are disjoint, and let $P_{G}$, $P_K$ and $P_{G|K}$ be
the marginal and conditional distributions induced by $P_{V}$,
respectively. Denote by $T_{\epsilon}^{(n)}(G)$ the projection of
$T_{\epsilon}^{(n)}(V)$ to the coordinates of $G$. Given any
$\textbf{k}\in\mathcal{K}^n$, denote $
T_{\epsilon}^{(n)}(G|\textbf{k})\triangleq
\left\{(G^n,\textbf{k})\in T_{\epsilon}^{(n)}(G, K)\right\}$.
Clearly $T_{\epsilon}^{(n)}(G|\textbf{k})=\emptyset$ if
$\textbf{k}\notin T_{\epsilon}^{(n)}(K)$. The following lemma (see,
e.g., \cite[pp. 342--343]{Cover_book}) restates the well known
exponential bounds for the cardinality of strongly typical sets.  In
the lemma $\eta=\eta(\epsilon,n)$ is a generic positive term such
that $\lim_{\epsilon\rightarrow
0}\lim_{n\rightarrow\infty}\eta(\epsilon,n)=0$.

\begin{lemma}\label{lemma1}
\rm{\cite{Cover_book}
\begin{enumerate}

\item For any $0<\epsilon_0<1$ we have
$P_{G|K}^{(n)}(T_{\epsilon}^{(n)}(G|\textbf{k})|\textbf{k})>
1-\epsilon_0$ for $n$ sufficiently large.

\item  $2^{n(H(K)-\eta)} \leq \left|\mathcal{T}_{\epsilon}^{(n)}(K)\right| \leq
2^{n(H(K)+\eta)}$.

\item For any $\textbf{k} \in \mathcal{T}_{\epsilon}^{(n)}(K)$, $2^{n(H(G|K)-\eta)} \leq
\left|\mathcal{T}_{\epsilon}^{(n)}(G|\textbf{k})\right| \leq
2^{n(H(G|K)+\eta)}$.

\end{enumerate}
}
\end{lemma}

Finally, we recall the Markov lemma for joint strong
$\epsilon$-typicality.

\begin{lemma} \label{lemma_MK}
{\rm (Markov lemma \cite[p. 579]{Cover_book}) Let $G \rightarrow K
\rightarrow L$ form a Markov chain in this order. For any
$0<\epsilon_0<1$ and $(\textbf{g},\textbf{k}) \in
T_{\epsilon}^{(n)}(G,K)$,
$$
P_{L|K}^{(n)}\left(\left.(\textbf{g},\textbf{k},L^n) \in
T_{\epsilon}^{(n)}(G,K,L)\right|\textbf{k}\right)>1-\epsilon_0
$$
for $n$ sufficiently large, independently of
$(\textbf{g},\textbf{k})$. }
\end{lemma}

\subsection{Outline of Proof}

It is enough to show that for given $Q_{U_1U_2}$, $W_{Y|X_1X_2}$, and
any $(R_1,R_2) \in \mathcal{R}_{in}(D_1,D_2)$, there exists a sequence
of codes $(f_1^{(n)}, f_2^{(n)}, \psi^{(n)})$ such that
$P_e^{(n)}\rightarrow 0$ as $n\rightarrow \infty$ and for any
$\delta>0$, 
\[
\frac{1}{n}\mathbb{E}[d_i(U^n_i,f_i^{(n)}(W_i,U^n_i))]
\leq D_i+\delta, \quad i=1,2
\]
for $n$ sufficiently large. Once this is
proved, a standard subsequence diagonalization argument can be used to
prove a similar statement with $\delta =0$, which then directly implies
the theorem.

Fix $(P_{T_1|U_1},P_{X_1|U_1T_1},P_{T_2|U_2},P_{X_2|U_2T_2})$ such
that $I(U_i;T_i)>0$ and the following are satisfied for some
$\epsilon'
>0$,
\begin{eqnarray}
&& R_1 < I(T_1;T_2,Y)-I(U_1;T_1)-\epsilon', \label{2pr_1}\\
&& R_2 < I(T_2;T_1,Y)-I(U_2;T_2)-\epsilon', \label{2pr_2} \\
&& R_1+R_2 < I(T_1,T_2;Y)-I(U_1,U_2;T_1,T_2)-\epsilon', \label{2pr_3}\\
&& \mathbb{E}[d_i(U_i,X_i)] \leq D_i, \hspace{1mm}i=1,2.
\label{2pr_4}
\end{eqnarray}
We will choose $f_1^{(n)}$ and $f_2^{(n)}$ in a random manner. For
$\epsilon<\frac{\delta}{2\max\{d_1^{max},d_2^{max}\}}$, define
$$
P^{(n)}_i\triangleq
\Pr\Bigl(\frac{1}{n}d_i\bigl(U^n_i,f_i^{(n)}(W_i,U^n_i)\bigr)>D_i+\epsilon
d_i^{max} \Bigr), \, i=1,2.
$$ 
The technically challenging part of the proof is to show that for any $0<\epsilon_1 \leq \frac{\delta}{6\max\{d_1^{max},d_2^{max}\}}$,
the probabilities $P_{e}^{(n)}$, $P_{1}^{(n)}$, and $P_{2}^{(n)}$,
when averaged over the random choice of $f_1^{(n)}$ and $f_2^{(n)}$,
satisfy
\begin{eqnarray}
\mathbb{E}[P_{e}^{(n)}]\leq \epsilon_1, \quad \mathbb{E}[P^{(n)}_1]
\leq \epsilon_1, \quad \mathbb{E}[P^{(n)}_2] \leq \epsilon_1
\nonumber
\end{eqnarray}
for $n$ sufficiently large. Then
$\mathbb{E}\{P_{e}^{(n)}+P^{(n)}_1+P^{(n)}_2\} \leq 3\epsilon_1$,
which guarantees that there exists at least one pair of codes
($f_1^{(n)}, f_2^{(n)}$) such that $
P_{e}^{(n)}+P^{(n)}_1+P^{(n)}_2\leq 3\epsilon_1$ and hence
$P_{e}^{(n)}\leq 3\epsilon_1$, $P^{(n)}_1\leq 3\epsilon_1$,
$P^{(n)}_2\leq 3\epsilon_1$ are simultaneously satisfied for $n$
sufficiently large. Finally, it can be easily shown that
$P^{(n)}_i\leq 3\epsilon_1$ implies for $n$ sufficiently large that
$$
\frac{1}{n}\mathbb{E}\left[d_i(U_i^n,f_i^{(n)}(W_i,U^n_i)\right]\leq
D_i+\epsilon d_i^{max} + P^{(n)}_i d_i^{max} \leq D_i+ \delta.
$$

\subsection{Random Code Design}

In what follows, the strongly $\epsilon$-typical set
$\mathcal{T}_{\epsilon}^{(n)}$ is defined under the joint
distribution
\begin{equation}
P_{U_1U_2T_1T_2X_1X_2Y}=Q_{U_1U_2}
P_{T_1|U_1}P_{X_1|U_1T_1}P_{T_2|U_2}P_{X_2|U_2T_2}W_{Y|X_1X_2}\label{jointd}
\end{equation}
and all the marginal and conditional distributions, e.g.,
$P_{U_2T_2}$, $P_{U_1|U_2T_2}$, etc, are induced by the joint
distribution. The parameter $\epsilon$, which is chosen to be
sufficiently small, will be specified in the proof.

\medskip

\textit{Generation of codebooks}. For $i=1,2$ and every
$w_i\in\mathcal{W}_i$, generate a codebook
$$
\mathcal{C}_{w_i}=\{\textbf{t}_i(w_i,1),\textbf{t}_i(w_i,2),...,\textbf{t}_i(w_i,L_i)\}
$$
with $L_i=2^{n[I(U_i;T_i)+4\epsilon]}$ codewords such that each
$\textbf{t}_i(w_i,l_i)$ is independently selected with uniform
distribution from the typical set
$\mathcal{T}_{\epsilon}^{(n)}(T_i)$. Denote the entire codebook for
Encoder $i$ by
$\mathcal{C}^{(i)}=\{\mathcal{C}_{w_i}\}_{w_i=1}^{M_i}$, where we
recall that $M_i=2^{nR_{i}}$. For each $\textbf{u}_i$ and codeword
$\textbf{t}_i(w_i,l_i)$ ($1\leq w_i\leq M_i, 1\leq l_i\leq L_i$),
generate a codeword $\textbf{x}_{i}$ according to
$P_{X_i|U_iT_i}^{(n)}(\textbf{x}_{i}|\textbf{u}_i,\textbf{t}_{i})$.
Denote the codebook of all the codewords $\textbf{x}_{i}$ by
$\mathcal{B}^{(i)}$.

\medskip

\textit{Encoder $f_1^{(n)}$}: Encoder $f_1^{(n)}$ is the
concatenation of a pre-encoder $\varphi_1^{(n)}:
\mathcal{W}_1\times\mathcal{U}_1^n\longrightarrow \mathcal{T}_1^n$
and a mapping
$g_1^{(n)}:\mathcal{U}_1^n\times\mathcal{T}_1^n\longrightarrow\mathcal{X}_1^n$.

To define $\varphi_1^{(n)}$, we need the following notation adopted
from \cite{Oohama97}. We introduce a conditional probability
$$
A^{(n)}(\textbf{u}_1,\textbf{t}_1)\triangleq
P_{U_2T_2|U_1T_1}^{(n)}\left(\left. (\textbf{u}_2,\textbf{t}_2):
(\textbf{u}_2,\textbf{t}_2)\in
\mathcal{T}_{\epsilon}^{(n)}(U_2T_2|\textbf{u}_1,\textbf{t}_1)\right|\textbf{u}_1,\textbf{t}_1\right).
$$
For $\mu \in(0,1)$, let
\begin{eqnarray}
\mathcal{F}^{(n)}_{\mu,\epsilon}(U_1,T_1)\triangleq\left\{(\textbf{u}_1,\textbf{t}_1):
A^{(n)}(\textbf{u}_1,\textbf{t}_1) \geq 1-\mu\right\}.\nonumber
\end{eqnarray}
By definition, we have $\mathcal{F}^{(n)}_{\mu,\epsilon}(U_1,T_1)
\subseteq \mathcal{T}_{\epsilon}^{(n)}(U_1,T_1)$.


We now describe the pre-encoding function
$\varphi_1^{(n)}=\varphi_1^{(n)}(w_1,\textbf{u}_1)$ which maps every
pair $(w_1,\textbf{u}_1)$ to a codeword in
$\mathcal{C}^{(1)}\subseteq\mathcal{T}_1^n$. Given
$w_1\in\{1,2,...,M_1\}$ and $\textbf{u}_1$, $\varphi_1^{(n)}$ seeks
the first codeword $\textbf{t}_1(w_1,l_1)$ (if any) in
$\mathcal{C}_{w_1}$ such that
$(\textbf{u}_1,\textbf{t}_1(w_1,l_1))\in
\mathcal{F}^{(n)}_{\mu,\epsilon}(U_1,T_1)$. If there is no such
codeword, $\varphi_1^{(n)}$ outputs $\textbf{t}_1(w_1,1)$. Next, for
each output $\textbf{t}_1(w_1,l_1)$ and $\textbf{u}_1$, $g_1^{(n)}$
sends out the associated codeword $\textbf{x}_1(w_1,\textbf{u}_1)$
to the channel. Thus,
$f_1^{(n)}(w_1,\textbf{u}_1)=g_1^{(n)}\left(\textbf{u}_1,\varphi_1^{(n)}(w_1,\textbf{u}_1)\right)$.

\medskip

\textit{Encoder $f_2^{(n)}$}: Encoder $f_2^{(n)}$ is the
concatenation of a pre-encoder $\varphi_2^{(n)}:
\mathcal{W}_2\times\mathcal{U}_2^n\longrightarrow \mathcal{T}_2^n$
and a mapping
$g_2^{(n)}:\mathcal{U}_2^n\times\mathcal{T}_2^n\longrightarrow\mathcal{X}_2^n$.

To define $\varphi_2^{(n)}$, let
\begin{eqnarray}
B^{(n)}_{\varphi_1}(\textbf{u}_2,\textbf{t}_2)&\triangleq&
\frac{1}{2^{nR_1}}\sum_{w_1=1}^{M_1}
P_{U_1|U_2T_2}^{(n)}\left(\left.\textbf{u}_1:
(\textbf{u}_1,\varphi_1^{(n)}(w_1,\textbf{u}_1))\in
\mathcal{T}_{\epsilon}^{(n)}(U_1T_1|\textbf{u}_2,\textbf{t}_2)\right|\textbf{u}_2,\textbf{t}_2\right).\nonumber
\end{eqnarray}
Also, for $\nu \in(0,1)$, define
$$
\mathcal{F}^{(n)}_{\varphi_1,\nu,\epsilon}(U_2,T_2)\triangleq\left\{(\textbf{u}_2,\textbf{t}_2):
B^{(n)}_{\varphi_1}(\textbf{u}_2,\textbf{t}_2) \geq 1-\nu\right\}.
$$
By definition, it is seen that
$\mathcal{F}^{(n)}_{\varphi_1,\nu,\epsilon}(U_2,T_2) \subseteq
\mathcal{T}_{\epsilon}^{(n)}(U_2,T_2)$.


We now describe the pre-encoding function
$\varphi_2^{(n)}=\varphi_2^{(n)}(w_2,\textbf{u}_2)$ which maps every
pair $(w_2,\textbf{u}_2)$ to a codeword in
$\mathcal{C}^{(2)}\subseteq\mathcal{T}_2^n$. Given
$w_2\in\{1,2,...,M_2\}$ and $\textbf{u}_2$, $\varphi_2^{(n)}$ seeks
the first codeword $\textbf{t}_2(w_2,l_2)$ (if any) in
$\mathcal{C}_{w_2}$ such that
$(\textbf{u}_2,\textbf{t}_2(w_2,l_2))\in
\mathcal{F}^{(n)}_{\varphi_1,\nu,\epsilon}(U_2,T_2)$. If there is no
such codeword, $\varphi_2^{(n)}$ outputs $\textbf{t}_2(w_2,1)$.
Next, for each output $\textbf{t}_2(w_2,l_2)$, $g_2^{(n)}$ sends out
the associated codeword $\textbf{x}_2(w_2,\textbf{u}_2)$ to the
channel. Thus,
$f_2^{(n)}(w_2,\textbf{u}_2)=g_2^{(n)}\left(\textbf{u}_2,\varphi_2^{(n)}(w_2,\textbf{u}_2)\right)$.

\medskip

\textit{Decoder $\psi^{(n)}$}: Given $\textbf{y}$, $\psi^{(n)}$
seeks
$\textbf{t}_1(\widehat{w}_1,\widehat{l}_1)\in\mathcal{C}^{(1)}$ and
$\textbf{t}_2(\widehat{w}_2,\widehat{l}_2)\in\mathcal{C}^{(2)}$ such
that
$$
(\textbf{t}_1(\widehat{w}_1,\widehat{l}_1),\textbf{t}_2(\widehat{w}_2,\widehat{l}_2),\textbf{y})\in
\mathcal{T}_{\epsilon}^{(n)}(T_1,T_2,Y).
$$ If such a pair
$(\textbf{t}_1(\widehat{w}_1,\widehat{l}_1),\textbf{t}_2(\widehat{w}_2,\widehat{l}_2))$
exists for a unique ($\widehat{w}_1,\widehat{w}_2$), then
$\psi^{(n)}$ outputs $\widehat{w}_1$ and $\widehat{w}_2$ as the
decoded messages. If there is no such pair
$(\widehat{w}_1,\widehat{w}_2)$, or it is not unique, a decoding
error is declared. Letting
$\textbf{t}_i(w_i,l_i)=\varphi_i^{(n)}(w_i,\textbf{u}_i)$, it is
easy to see that if there is a decoding error, then at least one of
the following events occurs:

\begin{enumerate}

\item $E_1$: $
(\textbf{t}_1(w_1,l_1),\textbf{t}_2(w_2,l_2),\textbf{y})\notin
\mathcal{T}_{\epsilon}^{(n)}(T_1,T_2,Y)$,

\item $E_2$: there exist $l'_1$ and $w'_1\neq w_1$ and $l'_2$ ($l'_2$
  may  or may not be equal to $l_2$) such that
$$
(\textbf{t}_1(w'_1,l'_1),\textbf{t}_2(w_2,l'_2),\textbf{y})\in
\mathcal{T}_{\epsilon}^{(n)}(T_1,T_2,Y),
$$

\item $E_3$: there exist $l'_2$ and $w'_2\neq w_2$ and $l'_1$ ($l'_1$
  may or may not be equal to $l_1$) such that
$$
(\textbf{t}_1(w_1,l'_1),\textbf{t}_2(w'_2,l'_2),\textbf{y})\in
\mathcal{T}_{\epsilon}^{(n)}(T_1,T_2,Y),
$$

or

\item $E_4$: there exist $l'_1$ and $w'_1\neq w_1$ and $l'_2$ and $w'_2\neq w_2$ such that
$$
(\textbf{t}_1(w'_1,l'_1),\textbf{t}_2(w'_2,l'_2),\textbf{y})\in
\mathcal{T}_{\epsilon}^{(n)}(T_1,T_2,Y).
$$
\end{enumerate}

In the following, we will bound the probabilities $P_{e}^{(n)}$,
$P^{(n)}_1$ and $P^{(n)}_2$ averaged over the random choice of all
codes $\mathcal{B}^{(1)}$, $\mathcal{B}^{(2)}$, $\mathcal{C}^{(1)}$,
and $\mathcal{C}^{(2)}$. To simplify the notation we abbreviate
$\mathbb{E}_{\mathcal{B}^{(1)},\mathcal{B}^{(2)},\mathcal{C}^{(1)},\mathcal{C}^{(2)}}[\;\cdot\;]$
as $\mathbb{E}_{\Omega}[\;\cdot\;]$.

\subsection{Bounding
$\mathbb{E}_{\Omega}[P_{e}^{(n)}]$}

To analyze the average probability of error, we need the following
lemmas.

\begin{lemma} \label{lemma_EMK}
{\rm For any $w_1\in\mathcal{W}_1$, $w_2\in\mathcal{W}_2$, and any
$\epsilon_0,\epsilon\in(0,1)$, one can choose $\mu,\nu\in (0,1)$
small
  enough such that
\begin{equation}
\mathbb{E}_{\mathcal{C}^{(1)},\mathcal{C}^{(2)}}\left[P_{U_1U_2}^{(n)}\left(
(\varphi_1^{(n)}(w_1,\textbf{u}_1), \textbf{u}_1, \textbf{u}_2,
\varphi_2^{(n)}(w_2,\textbf{u}_2))\in
\mathcal{T}^{(n)}_{\epsilon}(T_1,U_1,U_2,T_2) \right) \right] \geq
1- \epsilon_0\nonumber
\end{equation}
for $n$ sufficiently large, where the expectation is taken with
respect to the random codes $\mathcal{C}^{(1)}$ and
$\mathcal{C}^{(2)}$. }
\end{lemma}

The proof of Lemma \ref{lemma_EMK} is very similar to the proof of
the extended Markov lemma in \cite[Lemma 3]{Oohama97} for correlated
Gaussian sources and is hence omitted; readers may also refer to
\cite[Section 5.4.5]{yadongthesis}.

Since the watermarks are independently and uniformly distributed,
and by the symmetry of the code construction, we can assume without
the loss of generality that some fixed $w_1\in \mathcal{W}_1$ and
$w_2\in \mathcal{W}_2$ are the transmitted watermarks. Thus we bound
the probability of error as
\begin{eqnarray}
P^{(n)}_{e} &=&\Pr\left(\left\{\psi^{(n)}(Y^n)\neq
(w_1,w_2)\right\}\right)\nonumber\\
&\leq &
\Pr(A_1)+\left.\Pr\left(\left\{\psi^{(n)}(Y^n)\neq
(w_1,w_2)\right\}\right|A_1^c\right)\label{perror}
\end{eqnarray}
where $A_1$ is the event
$$
A_1: (\textbf{t}_1(w_1,l_1),\textbf{u}_1,\textbf{u}_2,
\textbf{t}_2(w_2,l_2),\textbf{x}_1,\textbf{x}_{2}) \notin
\mathcal{T}_{\epsilon}^{(n)}(T_1,U_1,U_2,T_2,X_1,X_2).
$$
Recall that
$\textbf{t}_i(w_i,l_i)=\varphi_i^{(n)}(w_i,\textbf{u}_i)$, $i=1,2$.
We also let $\textbf{t}_i(w_i,l_i')$ and $\textbf{t}_i(w'_i,l_i')$
be the $l_i'$-th codeword in the codebook $\mathcal{C}_{w_i}$ and
$\mathcal{C}_{w'_i}$, respectively.

We then introduce the event
$$
A_{0}: ( \textbf{t}_1(w_1,l_1),\textbf{u}_1, \textbf{u}_2,
\textbf{t}_2(w_2,l_2))\notin
\mathcal{T}_{\epsilon}^{(n)}(T_1,U_1,U_2,T_2).
$$
Taking expectation in (\ref{perror}) and using the union bound, we
have
\begin{equation}
\mathbb{E}_{\Omega} [P^{(n)}_{e} ]\leq \mathbb{E}_{\Omega}\Pr
\left(A_0\right)+\mathbb{E}_{\Omega}\Pr
\left(A_1|A_0^c\right)+\mathbb{E}_{\Omega}\Pr
\left(E_{1}|A_1^c\right)+\sum\limits_{k=2}^4
\mathbb{E}_{\Omega}\Pr \left(E_{k}|A_1^c\right).\label{pr_qw0}
\end{equation}

It immediately follows from Lemma \ref{lemma_EMK} that
\begin{equation}
\mathbb{E}_{\Omega}\Pr \left( A_0
\right)=\mathbb{E}_{\mathcal{C}^{(1)},\mathcal{C}^{(2)}}\Pr
\left( A_0 \right)\leq \epsilon_0 \label{probt1}
\end{equation}
for $n$ sufficiently large, where we set $\epsilon_0=\epsilon_1/7$
for a given $\epsilon_1\geq 0$ throughout the proof. When $A_0^c$
holds, since $\textbf{x}_1$ and $\textbf{x}_2$ are respectively
drawn according to the conditional probabilities
$P_{X_1|U_1T_1}^{(n)}(\cdot|\textbf{u}_1,\textbf{t}_1)$ and
$P_{X_2|U_2T_2}^{(n)}(\cdot|\textbf{u}_2,\textbf{t}_2)$, and
$\textbf{y}$ is drawn according to the conditional distribution
$W_{Y|X_1X_2}^{(n)}(\cdot|\textbf{x}_1,\textbf{x}_2)$, it follows
from two successive applications of Lemma \ref{lemma_MK} that
\begin{equation}
\mathbb{E}_{\Omega}\Pr \left(\left. A_1
\right|A_0^c\right)\leq \mathbb{E}_{\Omega}[\epsilon_0]=\epsilon_0
\label{probt2}
\end{equation}
and
\begin{eqnarray}
\lefteqn{\mathbb{E}_{\Omega}\Pr
\left(\left. E_1 \right|A_1^c\right)}\nonumber\\
&\leq
&\mathbb{E}_{\Omega}\Pr\left(\left.\left\{\left(\varphi_1^{(n)}(w_1,U^n_1),U^n_1,U^n_2,
\varphi_2^{(n)}(w_2,U^n_2),f_1^{(n)}(w_1,U^n_1),
f_2^{(n)}(w_2,U^n_2), Y^n\right) \notin
\mathcal{T}_{\epsilon}^{(n)}\right\}\right|A_1^c\right)\nonumber\\
&\leq  & \mathbb{E}_{\Omega}[\epsilon_0]= \epsilon_0 \label{probt3}
\end{eqnarray}
for $n$ sufficiently large. It remains to bound
$\mathbb{E}_{\Omega}\Pr \left\{\left. E_{k}\right|
A_1^c\right\}$ for $k=2,3,4$. Using the union bound we write
\begin{eqnarray}
\lefteqn{\mathbb{E}_{\Omega}\Pr \left(\left. E_{2}\right|
A_1^c\right)}\nonumber\\
&\leq &\sum_{w'_1\neq
w_1}\sum_{l_1'=1}^{L_{1}}\Pr\left(\left.\left\{(T^n_1(w'_1,l'_1),Y^n,T^n_2(w_2,l'_2))\in
\mathcal{T}_{\epsilon}^{(n)}(T_1,T_2,Y)\right\}\right|A_1^c\right),\label{bndE21}
\end{eqnarray}
where $T^n_1(w'_1,l'_1)$ is a RV uniformly drawn from
$\mathcal{T}_{\epsilon}^{(n)}(T_1)$ which is independent of
$(T^n_2(w_2,l'_2),Y^n)$ since $w'_1\neq w_1$. Thus we have
\begin{eqnarray}
\lefteqn{\Pr\left(\left.\left\{(T^n_1(w'_1,l'_1),Y^n,T^n_2(w_2,l'_2))\in
\mathcal{T}_{\epsilon}^{(n)}(T_1,T_2,Y)\right\}\right|A_1^c\right)}\nonumber\\
&=&\sum_{(\textbf{t}_2,\textbf{y})\in
\mathcal{T}_{\epsilon}^{(n)}(T_2,Y)}\sum_{\textbf{t}_1\in\mathcal{T}_{\epsilon}^{(n)}(T_1|\textbf{t}_2,\textbf{y})}\Pr\left(\left.T^n_2(w_2,l'_2)=\textbf{t}_2,Y^n=\textbf{y}\right|A_1^c\right)\nonumber\\
&&\qquad\qquad\Pr\left(\left.T^n_1(w'_1,l'_1)=\textbf{t}_1\right|T^n_2(w_2,l'_2)=\textbf{t}_2,Y^n=\textbf{y},A_1^c\right)\nonumber\\
&=& \sum_{(\textbf{t}_2,\textbf{y})\in
\mathcal{T}_{\epsilon}^{(n)}(T_2,Y)}\sum_{\textbf{t}_1\in\mathcal{T}_{\epsilon}^{(n)}(T_1|\textbf{t}_2,\textbf{y})}\Pr\left(\left.T^n_2(w_2,l'_2)=\textbf{t}_2,Y^n=\textbf{y}\right|A_1^c\right)\mbox{Pi}\left(T^n_1(w'_1,l'_1)=\textbf{t}_1\right)\nonumber\\
&=& \sum_{(\textbf{t}_2,\textbf{y})\in
\mathcal{T}_{\epsilon}^{(n)}(T_2,Y)}\Pr\left(\left.T^n_2(w_2,l_2)=\textbf{t}_2,Y^n=\textbf{y}\right|A_1^c\right)\frac{|\mathcal{T}_{\epsilon}^{(n)}(T_1|\textbf{t}_2,\textbf{y})|}{|\mathcal{T}_{\epsilon}^{(n)}(T_1)|}\nonumber\\
&\leq &
\frac{2^{n[H(T_1|T_2,Y)+\eta]}}{2^{n[H(T_1)-\eta]}}\sum_{(\textbf{t}_2,\textbf{y})\in
\mathcal{T}_{\epsilon}^{(n)}(T_2,Y)}\Pr\left(\left.T^n_2(w_2,l'_2)=\textbf{t}_2,Y^n=\textbf{y}\right|A_1^c\right)\nonumber\\
&\leq & 2^{-n[I(T_1;T_2,Y)-2\eta]},\label{bndE1}
\end{eqnarray}
where the first inequality follows from Lemma \ref{lemma1}.
Recalling that $\eta\rightarrow 0$ as $n\rightarrow \infty$ and
$\epsilon\rightarrow 0$, we can make sure that
$2\eta<\epsilon'-4\epsilon$ by choosing $\epsilon$ small enough and
$n$ large enough. Thus from (\ref{bndE21})
\begin{eqnarray}
\mathbb{E}_{\Omega}\Pr \left(\left. E_{2}\right|
A_1^c\right)&\leq&
2^{n[R_1+I(U_1;T_1)+4\epsilon-I(T_1;T_2,Y)+2\eta]}\nonumber\\
&\leq&
2^{n[R_1+I(U_1;T_1)-I(T_1;T_2,Y)+\epsilon']}\nonumber\\
&\leq& \epsilon_0\label{probt4}
\end{eqnarray}
for $\epsilon$ sufficiently small and $n$ sufficiently large, where
(\ref{probt4}) follows from the assumption (\ref{2pr_1}). Similarly
we have
\begin{equation}
\mathbb{E}_{\Omega}\Pr \left(\left.
E_{3}\right|A_1^c\right)\leq \epsilon_0 \label{probt5}
\end{equation}
for $\epsilon$ small enough and $n$ sufficiently large. We next
bound
\begin{eqnarray}
\lefteqn{\mathbb{E}_{\Omega}\Pr
\left(\left. E_{4}\right|A_1^c\right)}\nonumber\\
&\leq & \sum_{w'_1\neq w_1}\sum_{l_1'=1}^{L_{1}}\sum_{w'_2\neq
w_2}\sum_{l_2'=1}^{L_{2}}\Pr\left(\left.\left\{(T_1^n(w'_1,l'_1),T_2^n(w'_2,l'_2),Y^n)\in
\mathcal{T}_{\epsilon}^{(n)}(T_1,T_2,Y)\right\}\right|A_1^c\right),\nonumber
\end{eqnarray}
where $T^n_1(w'_1,l'_1)$ and $T^n_2(w'_2,l'_2)$ are RVs
independently drawn from $\mathcal{T}_{\epsilon}^{(n)}(T_1)$ and
$\mathcal{T}_{\epsilon}^{(n)}(T_2)$ according to the uniform
distribution, respectively. We have
\begin{eqnarray}
\lefteqn{\Pr\left(\left.\left\{(T_1^n(w'_1,l'_1),T_2^n(w'_2,l'_2),Y^n)\in
\mathcal{T}_{\epsilon}^{(n)}(T_1,T_2,Y)\right\}\right|A_1^c\right)}\nonumber\\
&=&
\sum_{\textbf{y}\in\mathcal{T}_{\epsilon}^{(n)}(Y)}\sum_{(\textbf{t}_1,\textbf{t}_2)\in
\mathcal{T}_{\epsilon}^{(n)}(T_1,T_2|\textbf{y})}\Pr(Y^n=\textbf{y}|A_1^c)\nonumber\\
&&
\qquad\qquad\Pr(T_1^n(w'_1,l'_1)=\textbf{t}_1,T_2^n(w'_2,l'_2)=\textbf{t}_2|A_1^c,Y^n=\textbf{y})\nonumber\\
&=&
\sum_{\textbf{y}\in\mathcal{T}_{\epsilon}^{(n)}(Y)}\sum_{(\textbf{t}_1,\textbf{t}_2)\in
\mathcal{T}_{\epsilon}^{(n)}(T_1,T_2|Y)}\Pr(Y^n=\textbf{y}|A_1^c)\frac{1}{|\mathcal{T}_{\epsilon}^{(n)}(T_1)|}\frac{1}{|\mathcal{T}_{\epsilon}^{(n)}(T_2)|}\nonumber\\
&\leq&\sum_{\textbf{y}\in\mathcal{T}_{\epsilon}^{(n)}(Y)}\Pr(Y^n=\textbf{y}|A_1^c)\frac{2^{n[H(T_1,T_2|Y)+\eta]}}{2^{n[H(T_1)-\eta]}2^{n[H(T_2)-\eta]}}\nonumber\\
&\leq & 2^{-n[I(T_1,T_2;Y)+I(T_1;T_2)-3\eta]}\nonumber
\end{eqnarray}
and hence
\begin{eqnarray}
\lefteqn{\mathbb{E}_{\Omega}\Pr
\left(\left. E_{4}\right|A_1^c\right)}\nonumber\\
&\leq & 2^{n[R_1+R_2+I(U_1;T_1)+I(U_2;T_2)-I(T_1,T_2;Y)-I(T_1;T_2)+8\epsilon+3\eta]}\nonumber\\
&\leq &
2^{n[R_1+I(U_1,U_2;T_1,T_2)-I(T_1,T_2;Y)+\epsilon']}\nonumber\\
&\leq& \epsilon_0 \label{probt8}
\end{eqnarray}
for $n$ sufficiently large and $\epsilon$ small enough (such that
$8\epsilon+3\eta<\epsilon'$), where the second inequality holds by the
Markov chain relation $T_1\rightarrow U_1\rightarrow U_2\rightarrow
T_2$ imposed in Definition~\ref{def2},  and the last inequality follows
from the assumption (\ref{2pr_3}). Finally, substituting
(\ref{probt1})--(\ref{probt3}), (\ref{probt4}), (\ref{probt5}) and
(\ref{probt8}) into (\ref{pr_qw0}) yields $\mathbb{E}_{\Omega}
[P^{(n)}_{e} ]\leq 7\epsilon_0=\epsilon_1$ for $\epsilon$ sufficiently
small and $n$ sufficiently large.

\subsection{Bounding
$\mathbb{E}_{\Omega}[P^{(n)}_i]$}

We only bound $\mathbb{E}_{\Omega} [P_i^{(n)}]$ for $i=1$, since the
case $i=2$ can be dealt with similarly. When $\left(\textbf{u}_1,
\textbf{x}_1(w_1,\textbf{u}_1)\right) \in
\mathcal{T}^{(n)}_{\epsilon}(U_1,X_1)$,
$$
\frac{1}{n}d_1\bigl(\textbf{u}_1,\textbf{x}_1(w_1,\textbf{u}_1)\bigr)\leq
\mathbb{E}[d_1(U_1,X_1)]+\epsilon d_1^{max} \leq D_1+\epsilon
d_1^{max}
$$ for $n$ sufficiently large, where the
first inequality follows from the definition of strong typicality
and the second inequality follows from (\ref{2pr_4}). This means
that if $\frac{1}{n}
d_1\bigl(U^n_1,f_1^{(n)}(W_1,U_1^n)\bigr)>D_1+\epsilon d_1^{max}$,
then we must have $\bigl(U_1^n, f_1^{(n)}(W_1,U_1^n)\bigr) \notin
\mathcal{T}^{(n)}_{\epsilon}(U_1,X_1)$ for $n$ sufficiently large.
Thus, we can bound
\begin{eqnarray}
\lefteqn{\Pr\Bigl( \frac{1}{n}
d_1(U^n_1,f_1^{(n)}(W_1,U_1^n))>D_1+\epsilon d_1^{max}  \Bigr)}
\nonumber\\
&\leq & \Pr\left(\bigl(U_1^n, f_1^{(n)}(W_1,U_1^n)\bigr)
\notin \mathcal{T}^{(n)}_{\epsilon}(U_1,X_1)\right)\nonumber\\
&\leq & \Pr\left(\bigl(U_1^n,
\varphi_1^{(n)}(W_1,U_1^n),f_1^{(n)}(W_1,U_1^n)\bigr) \notin
\mathcal{T}^{(n)}_{\epsilon}(U_1,T_1,X_1)\right)\nonumber\\
&\leq & \Pr\left(\bigl(U_1^n, \varphi_1^{(n)}(W_1,U_1^n)\bigr)
\notin
\mathcal{T}^{(n)}_{\epsilon}(U_1,T_1)\right)\nonumber\\
&&+\left.\Pr\left(\bigl(U_1^n,\varphi_1^{(n)}(W_1,U_1^n),
f_1^{(n)}(W_1,U_1^n)\bigr) \notin
\mathcal{T}^{(n)}_{\epsilon}(U_1,T_1,X_1)\right|(U_1^n,\varphi_1^{(n)}(W_1,U_1^n))\in
\mathcal{T}^{(n)}_{\epsilon}(U_1,T_1)\right)\nonumber\\
&\leq & \Pr\left( \bigl(\varphi_1^{(n)}(W_1,U_1^n), U_1^n,
U_2^n, \varphi_2^{(n)}(W_2,U_2^n)\bigr)\notin
\mathcal{T}^{(n)}_{\epsilon}(T_1,U_1,U_2,T_2) \right)\nonumber\\
&&+\left.\Pr\left(\bigl(U_1^n,\varphi_1^{(n)}(W_1,U_1^n),
f_1^{(n)}(W_1,U_1^n)\bigr) \notin
\mathcal{T}^{(n)}_{\epsilon}(U_1,T_1,X_1)\right|(U_1^n,\varphi_1^{(n)}(W_1,U_1^n))\in
\mathcal{T}^{(n)}_{\epsilon}(U_1,T_1)\right).\nonumber\\\label{disbnd}
\end{eqnarray}
Now taking expectation on both sides, the first term of
(\ref{disbnd}) is bounded by $\frac{\epsilon_1}{2}$ by Lemma
\ref{lemma_EMK}, and the second term is bounded by
$\frac{\epsilon_1}{2}$ for sufficiently large $n$ by Lemma
\ref{lemma1}. This completes the proof of the bound
$\mathbb{E}_{\Omega} [P_1^{(n)}]\leq \epsilon_1$ for $n$
sufficiently large. \qed

\section{Concluding Remarks}\label{secCon}

We have studied a multi-user information embedding system consisting
of two information embedders and one joint decoder connected via a
multiple-access attack channel.  We have obtained an inner bound for
the capacity region in a computable single-letter form. We also
derived an outer bound for the capacity region, but in this case  the
auxiliary random 
variables involved in the region's characterization have no upper bounds on
their alphabet's cardinality. Consequently, there may not exist an
algorithm to compute the outer bound with arbitrary precision.  We
have also addressed the special case when the covertexts are independent
of each other and inner and outer bounds for the capacity region of
this simplified system are provided.  Finally, we remark that using a
similar technique inner and outer bounds are derived in \cite[Chapter
  5]{yadongthesis} for the capacity region of private multi-user
embedding systems with quantization.

\section*{Appendix}

\appendix

\section{Proof of Theorem \ref{theo2}}\label{proof2}

The proof is a generalization of the proof of the converse in
\cite{Willems} for a single-user embedding system.

We need to show that any MAE code $(f_1^{(n)}, f_2^{(n)}, \psi^{(n)})$
with achievable rate pair $(R_1,R_2)$ must satisfy
(\ref{TH1_F1})--(\ref{TH1_F3}) for some auxiliary RVs $T_1$ and $T_2$
with joint distribution $P_{U_1U_2T_1T_2X_1X_2Y}\in
\mathcal{P}_{D_1,D_2}$. It follows from Fano's inequality that
$$
H(W_1,W_2|Y^n)\leq n(R_1+R_2)P_e^{(n)}+H(P_e^{(n)})\triangleq
n\epsilon_n.
$$
It is clear that $\epsilon_n\rightarrow 0$ if $P_e^{(n)}\rightarrow
0$ and
\begin{eqnarray}
H(W_1|Y^n)&\leq &H(W_1,W_2|Y^n)\leq n\epsilon_n,\nonumber\\
H(W_2|Y^n)&\leq &H(W_1,W_2|Y^n)\leq n\epsilon_n\nonumber.
\end{eqnarray}
Because $W_1$ is uniformly drawn from the message set
$\{1,2,...,2^{nR_1}\}$ and is independent of $U_1^n$, we have
$$
nR_1=H(W_1)=I(W_1;Y^n)+H(W_1|Y^n)\leq
I(W_1;Y^n)-\underbrace{I(W_1;U_1^n)}_{=0}+n\epsilon_n.
$$
Hence we can write
\begin{eqnarray}
\lefteqn{I(W_1;Y^n)-I(W_1;U_1^n)}\nonumber\\
&\stackrel{(a)}{=}&
\sum_{k=1}^{n}\left[I(W_1;Y_k|Y_1^{k-1})-I(W_1;U_{1k}|U_{1,k+1}^{n})\right]\nonumber\\
&=& \sum_{k=1}^{n}\left[H(Y_k|Y_1^{k-1})-H(Y_k|W_1,Y_1^{k-1},U_{1,k+1}^{n})-I(Y_k;U_{1,k+1}^{n}|W_1,Y_1^{k-1})\right.\nonumber\\
&&\quad\quad\left.-H(U_{1k}|U_{1,k+1}^{n})+H(U_{1k}|W_1,U_{1,k+1}^{n})\right]\nonumber\\
&\stackrel{(b)}{=}& \sum_{k=1}^{n}\left[H(Y_k|Y_1^{k-1})-H(Y_k|W_1,Y_1^{k-1},U_{1,k+1}^{n})-I(U_{1k};Y_1^{k-1}|W_1,U_{1,k+1}^{n})\right.\nonumber\\
&&\quad\quad\left.-H(U_{1k}|U_{1,k+1}^{n})+H(U_{1k}|W_1,U_{1,k+1}^{n})\right]\nonumber\\
&\stackrel{(c)}{=}& \sum_{k=1}^{n}\left[H(Y_k|Y_1^{k-1})-H(Y_k|W_1,Y_1^{k-1},U_{1,k+1}^{n})\right.\nonumber\\
&&\quad\quad\left.-H(U_{1k})+H(U_{1k}|W_1,Y_1^{k-1},U_{1,k+1}^{n})\right]\nonumber\\
&\leq &\sum_{k=1}^{n}\left[H(Y_k)-H(Y_k|W_1,U_{1,k+1}^{n},Y_1^{k-1})-I(U_{1k};W_1,Y_1^{k-1},U_{1,k+1}^{n})\right]\nonumber\\
&=& \sum_{k=1}^{n}\left[I(Y_k;W_1,U_{1,k+1}^{n},Y_1^{k-1})-I(U_{1k};W_1,Y_1^{k-1},U_{1,k+1}^{n})\right]\nonumber\\
&\stackrel{(d)}{\leq} & \sum_{k=1}^{n}\left[I(W_2,U_{2,k+1}^{n},Y_1^{k-1},Y_k;W_1,U_{1,k+1}^{n},Y_1^{k-1})-I(U_{1k};W_1,Y_1^{k-1},U_{1,k+1}^{n})\right]\nonumber\\
&\stackrel{(e)}{=}& \sum_{k=1}^n
[I(L_{2k},Y_k;L_{1k})-I(U_{1k};L_{1k})]\nonumber
\end{eqnarray}
where in (a) $Y_1^{k-1}\triangleq (Y_1,Y_2,...,Y_{k-1})$ and
$U_{1,k+1}^{n}\triangleq (U_{1,k+1},U_{1,k+2},...,U_{1,n})$, (b)
follows from the ``summation by parts'' identity \cite[Lemma
7]{CsiszarKorner}, (c) holds since the source $U_1$ is memoryless,
in (d) $U_{2,k+1}^{n}\triangleq (U_{2,k+1},U_{2,k+2},...,U_{2,n})$,
and in (e) $L_{1k}\triangleq (W_1,Y_1^{k-1},U_{1,k+1}^{n})$ and
$L_{2k}\triangleq (W_2,Y_1^{k-1},U_{2,k+1}^{n})$. Hence we obtain
the bound
\begin{equation}
R_1\leq \frac{1}{n}\sum_{k=1}^n
[I(L_{1k};L_{2k},Y_k)-I(U_{1k};L_{1k})]+\epsilon_n.\label{bndrate1}
\end{equation}
Similarly, we can show that
\begin{equation}
R_2\leq \frac{1}{n}\sum_{k=1}^n
[I(L_{2k};L_{1k},Y_k)-I(U_{2k};L_{2k})]+\epsilon_n.\label{bndrate2}
\end{equation}
To bound the sum of the rates, we write
\begin{eqnarray}
n(R_1+R_2)&=&H(W_1,W_2)=I(W_1,W_2;Y^n)+H(W_1,W_2|Y^n)\nonumber\\
&\leq &
I(W_1,W_2;Y^n)-\underbrace{I(W_1,W_2;U_1^n,U_2^n)}_{=0}+n\epsilon_n
\end{eqnarray}
and
\begin{eqnarray}
\lefteqn{I(W_1,W_2;Y^n)-I(W_1,W_2;U_1^n,U_2^n)}\nonumber\\
&=&
\sum_{k=1}^{n}\left[I(W_1,W_2;Y_k|Y_1^{k-1})-I(W_1,W_2;U_{1k},U_{2k}|U_{1,k+1}^{n},U_{2,k+1}^{n})\right]\nonumber\\
&=& \sum_{k=1}^{n}\left[H(Y_k|Y_1^{k-1})-H(Y_k|W_1,U_{1,k+1}^{n},Y_1^{k-1},W_2,U_{2,k+1}^{n})-I(Y_k;U_{1,k+1}^{n},U_{2,k+1}^{n}|W_1,W_2,Y_1^{k-1})\right.\nonumber\\
&&\quad\quad\left.-H(U_{1k},U_{2k}|U_{1,k+1}^{n},U_{2,k+1}^{n})+H(U_{1k},U_{2k}|W_1,W_2,U_{1,k+1}^{n},U_{2,k+1}^{n})\right]\nonumber\\
&=& \sum_{k=1}^{n}\left[H(Y_k|Y_1^{k-1})-H(Y_k|W_1,U_{1,k+1}^{n},Y_1^{k-1},W_2,U_{2,k+1}^{n})-I(U_{1k},U_{2k};Y_1^{k-1}|W_1,W_2,U_{1,k+1}^{n},U_{2,k+1}^{n})\right.\nonumber\\
&&\quad\quad\left.-H(U_{1k},U_{2k})+H(U_{1k},U_{2k}|W_1,W_2,U_{1,k+1}^{n},U_{2,k+1}^{n})\right]\nonumber\\
&=& \sum_{k=1}^{n}\left[H(Y_k|Y_1^{k-1})-H(Y_k|W_1,U_{1,k+1}^{n},Y_1^{k-1},W_2,U_{2,k+1}^{n})\right.\nonumber\\
&&\quad\quad\left.-H(U_{1k},U_{2k})+H(U_{1k},U_{2k}|W_1,W_2,U_{1,k+1}^{n},U_{2,k+1}^{n},Y_1^{k-1})\right]\nonumber\\
&\leq &\sum_{k=1}^{n}\left[H(Y_k)-H(Y_k|W_1,U_{1,k+1}^{n},W_2,U_{2,k+1}^{n},Y_1^{k-1})-I(U_{1k},U_{2k};L_{1k},L_{2k})\right]\nonumber\\
&=& \sum_{k=1}^{n}\left[I(Y_k;W_1,U_{1,k+1}^{n},W_2,U_{2,k+1}^{n},Y_1^{k-1})-I(U_{1k},U_{2k};L_{1k},L_{2k})\right]\nonumber\\
&=& \sum_{k=1}^n
[I(Y_k;L_{1k},L_{2k})-I(U_{1k},U_{2k};L_{1k},L_{2k})],\nonumber
\end{eqnarray}
which implies
\begin{equation}
R_1+R_2\leq
\frac{1}{n}\sum_{k=1}^{n}\left[I(L_{1k},L_{2k};Y_k)-I(U_{1k},U_{2k};L_{1k},L_{2k})\right]+\epsilon_n\label{bndrate3}.
\end{equation}

We next introduce a time-sharing RV to simplify the bounds
(\ref{bndrate1}), (\ref{bndrate2}), and (\ref{bndrate3}) using a
single-letter characterization. Define a RV $V$ with alphabet
$\{1,2,...,n\}$ and distribution $P_{V}(v)=1/n$. We next introduce
RVs $U_1$ and $U_2$ such that
$$
\Pr(U_1=u_1,U_2=u_2)=\Pr(U_{1k}=u_1,U_{2k}=u_2)=Q_{U_1U_2}(u_1,u_2)
$$
for all $(u_1,u_2)\in\mathcal{U}_1\times\mathcal{U}_2$, which are
independent of $V$. Furthermore, we define new RVs $L_1$, $L_2$,
$X_1$, $X_2$, and $Y$ by
\begin{eqnarray}
\lefteqn{\Pr(L_1=l_1, L_2=l_2, X_1=x_1, X_2=x_2,
Y=y|V=k)}\nonumber\\
&=&\Pr(L_{1k}=l_1, L_{2k}=l_2, X_{1k}=x_1, X_{2k}=x_2,
Y_k=y)\nonumber
\end{eqnarray}
for all
$(l_1,l_2,x_1,x_2,y)\in\mathcal{L}_1\times\mathcal{L}_2\times\mathcal{X}_1\times\mathcal{X}_2\times\mathcal{Y}$.
It follows that
\begin{eqnarray}
\lefteqn{\frac{1}{n}\sum_{k=1}^n
[I(L_{1k};L_{2k},Y_k)-I(U_{1k};L_{1k})]}\nonumber\\
&=& I(L_{1};L_{2},Y|V)-I(U_{1};L_{1}|V)\nonumber\\
&= &
H(L_1|V)-H(L_{1}|L_{2},Y,V)-H(U_1|V)+H(U_1|L_1,V)\nonumber\\
&\stackrel{(a)}{\leq} & H(L_1)-H(L_{1}|L_{2},Y,V)-H(U_1)+H(U_1|L_1,V)\nonumber\\
&=& I(L_{1};L_{2},Y,V)-I(U_1;L_1,V)\nonumber\\
&\leq & I(L_{1},V;L_{2},Y,V)-I(U_1;L_1,V)\nonumber\\
&\stackrel{(b)}{=}& I(T_1;T_{2},Y)-I(T_1;U_1)\nonumber
\end{eqnarray}
where (a) holds since conditioning reduces entropy and $U_1$ is
independent of $V$, and in (b) $T_1\triangleq (L_1,V)$ and
$T_2\triangleq (L_2,V)$. This shows that
\begin{equation}
R_1\leq I(T_1;T_{2},Y)-I(T_1;U_1)+\epsilon_n.\label{bndrate11}
\end{equation}
By a similar argument, we can show
\begin{equation}
R_2\leq I(T_2;T_{1},Y)-I(T_2;U_2)+\epsilon_n\label{bndrate21}
\end{equation}
and
\begin{equation}
R_1+R_2\leq
I(T_1,T_{2};Y)-I(U_1,U_2;T_1,T_2)+\epsilon_n\label{bndrate31}.
\end{equation}
For such RVs $(U_1,U_2,T_1,T_2,X_1,X_2,Y)$, it can be readily seen
that the Markov chain relation $(U_1,U_2,T_1,T_2)\rightarrow
(X_1,X_2)\rightarrow Y$ holds. In fact,
\begin{eqnarray}
\lefteqn{\Pr(Y=y|U_1=u_1,U_2=u_2,T_1=t_1=(l_1,k),T_2=t_2=(l_2,k),X_1=x_1,X_2=x_2)}\nonumber\\
&=&\Pr(Y=y|U_1=u_1,U_2=u_2,L_1=l_1,L_2=l_2,X_1=x_1,X_2=x_2,V=k)\nonumber\\
&=&\Pr(Y_k=y|U_{1k}=u_1,U_{2k}=u_2,L_{1k}=l_1,L_{2k}=l_2,X_{1k}=x_1,X_{2k}=x_2)\nonumber\\
&=&\Pr(Y_k=y|X_{1k}=x_1,X_{2k}=x_2)\nonumber\\
&=&W_{Y|X_1X_2}(y|x_1,x_2).\nonumber
\end{eqnarray}
Next we bound the distortions $\mathbb{E}[d_i(U_i,X_i)]$. Since
$(R_1,R_2)$ is achievable under the sequence of codes $(f_1^{(n)},
f_2^{(n)}, \psi^{(n)})$, this implies that for any $\delta>0$ and all
$n$ large enough, we have
\begin{eqnarray}
D_i+\delta &\geq &\frac{1}{n}\frac{1}{2^{nR_i}}\sum_{w_i=1}^{M_i}\sum_{\mathcal{U}_i^n}Q_{U_i}^{(n)}(\textbf{u}_i)d_i\left(\textbf{u}_i,f_i^{(n)}(w_i,\textbf{u}_i)\right)\nonumber\\
&=&\frac{1}{n}\sum_{\mathcal{U}_i^n\times\mathcal{X}_i^n}\Pr(U^n_i=\textbf{u}_i,X^n_i=\textbf{x}_i)d_i(\textbf{u}_i,\textbf{x}_i)\nonumber\\
&=&\frac{1}{n}\sum_{k=1}^{n}\sum_{\mathcal{U}_i^n\times\mathcal{X}_i^n}\Pr(U^n_i=\textbf{u}_i,X^n_i=\textbf{x}_i)d_{i}(u_{ik},x_{ik})\nonumber\\
&=& \sum_{k=1}^{n}P_V(V=k)\sum_{\mathcal{U}_i\times\mathcal{X}_i}\Pr(U_{ik}=u_{ik},X_{ik}=x_{ik})d_{i}(u_{ik},x_{ik})\nonumber\\
&=& \sum_{k=1}^{n}P_V(V=k)\sum_{\mathcal{U}_i\times\mathcal{X}_i}\Pr(U_{i}=u_{i},X_{i}=x_{i}|V=k)d_{i}(u_{i},x_{i})\nonumber\\
&=& \sum_{k=1}^{n}\sum_{\mathcal{U}_i\times\mathcal{X}_i}\Pr(U_{i}=u_{i},X_{i}=x_{i},V=k)d_{i}(u_{i},x_{i})\nonumber\\
&=&\sum_{\mathcal{U}_i\times\mathcal{X}_i}P_{U_iX_i}(u_i,x_i)d_i(u_i,x_i).\nonumber
\end{eqnarray}

Thus we obtained that $\mathbb{E}[d_i(U_i,X_i)]\le D_i+\delta$ for
$i=1,2$. Combined with (\ref{bndrate11})--(\ref{bndrate31}) and
recalling that $\lim_{n\to \infty} \epsilon_n =0$ and that
$\mathcal{R}(D_1,D_2)$ is closed, we conclude that $\mathcal{R}(D_1,D_2)
\subset\mathcal{R}_{out}(D_1+\delta,D_2+\delta)$ as claimed. \qed

\section{Proof of Theorem \ref{theo3}}\label{proof3}

The forward part (achievability) is a consequence of Theorem
\ref{theo1} since $(U_1,T_1)$ and $(U_2,T_2)$ are independent and
hence $I(T_1;T_2,Y)=I(T_1;Y|T_2)$, $I(T_2;T_1,Y)=I(T_2;Y|T_1)$, and
$I(U_1,U_2;T_1,T_2)=I(U_1;T_1)+I(U_2;T_2)$. To prove the converse
part, we need to sharpen the bounds in the last proof. We start from
\begin{eqnarray}
\lefteqn{I(W_1;Y^n)-I(W_1;U_1^n)}\nonumber\\
&=&\sum_{k=1}^{n}\left[I(Y_k;W_1,U_{1,k+1}^{n}|Y_1^{k-1})-I(U_{1k};W_1,Y_1^{k-1},U_{1,k+1}^{n})\right]\nonumber\\
&=&\sum_{k=1}^{n}\left[H(W_1,U_{1,k+1}^{n}|Y_1^{k-1})-H(W_1,U_{1,k+1}^{n}|Y_1^{k-1},Y_k)-I(U_{1k};W_1,Y_1^{k-1},U_{1,k+1}^{n})\right]\nonumber\\
&\stackrel{(a)}{= }&
\sum_{k=1}^{n}\left[H(W_1,U_{1,k+1}^{n}|W_2,U_{2,k+1}^{n},Y_1^{k-1})-H(W_1,U_{1,k+1}^{n}|W_2,U_{2,k+1}^{n},Y_1^{k-1},Y_k)\right.\nonumber\\
&&\qquad\qquad\qquad\left.-I(U_{1k};W_1,Y_1^{k-1},U_{1,k+1}^{n})\right]\nonumber\\
&=&
\sum_{k=1}^{n}\left[I(W_1,U_{1,k+1}^{n};Y_k|W_2,U_{2,k+1}^{n},Y_1^{k-1})-I(U_{1k};W_1,Y_1^{k-1},U_{1,k+1}^{n})\right]\nonumber\\
&\leq & \sum_{k=1}^{n}\left[I(W_1,U_{1,k+1}^{n},Y_1^{k-1};Y_k|W_2,U_{2,k+1}^{n},Y_1^{k-1})-I(U_{1k};W_1,Y_1^{k-1},U_{1,k+1}^{n})\right]\nonumber\\
&=& \sum_{k=1}^n [I(L_{1k};Y_k|L_{2k})-I(U_{1k};L_{1k})]\nonumber
\end{eqnarray}
where (a) follows since $(W_1,U_{1,k+1}^{n})$  is now independent of
$(W_2,U_{2,k+1}^{n})$, and in the last equality we still let
$L_{1k}\triangleq (W_1,Y_1^{k-1},U_{1,k+1}^{n})$ and
$L_{2k}\triangleq (W_2,Y_1^{k-1},U_{2,k+1}^{n})$. Thus, using Fano's
inequality we have
$$
R_1\leq \frac{1}{n}\sum_{k=1}^n
[I(L_{1k};Y_k|L_{2k})-I(U_{1k};L_{1k})]+\epsilon_n.
$$
Similarly we can obtain
$$
R_2\leq \frac{1}{n}\sum_{k=1}^n
[I(L_{2k};Y_k|L_{1k})-I(U_{2k};L_{2k})]+\epsilon_n.
$$
To bound the sum of the rates, we have
\begin{eqnarray}
n(R_1+R_2)&=&H(W_1,W_2)=I(W_1,W_2;Y^n)+H(W_1,W_2|Y^n)\nonumber\\
&\leq & I(W_1,W_2;Y^n)-I(W_1;U_1^n)-I(W_2;U_2^n)+n\epsilon_n
\end{eqnarray}
and
\begin{eqnarray}
\lefteqn{I(W_1,W_2;Y^n)-I(W_1;U_1^n)-I(W_2;U_2^n)}\nonumber\\
&=&
\sum_{k=1}^{n}\left[I(W_1;Y_k|Y_1^{k-1})+I(W_2;Y_k|W_1,Y_1^{k-1})-I(W_1;U_{1k}|U_{1,k+1}^{n})-I(W_2;U_{2k}|U_{2,k+1}^{n})\right]\nonumber\\
&= & \sum_{k=1}^{n}\left[H(Y_k|Y_1^{k-1})-H(Y_k|W_1,Y_1^{k-1},U_{1,k+1}^{n})-I(Y_k;U_{1,k+1}^{n}|W_1,Y_1^{k-1})\right.\nonumber\\
&&\quad\quad+H(Y_k|W_1,Y_1^{k-1})-H(Y_k|W_1,W_2,Y_1^{k-1},U_{2,k+1}^{n})-I(Y_k;U_{2,k+1}^{n}|W_1,W_2,Y_1^{k-1})\nonumber\\
&&\quad\quad\left.-H(U_{1k}|U_{1,k+1}^{n})+H(U_{1k}|W_1,U_{1,k+1}^{n})-H(U_{2k}|U_{2,k+1}^{n})+H(U_{2k}|W_2,U_{2,k+1}^{n})\right]\nonumber\\
&= & \sum_{k=1}^{n}\left[H(Y_k|Y_1^{k-1})-H(Y_k|W_1,Y_1^{k-1},U_{1,k+1}^{n})-I(U_{1k};Y_1^{k-1}|W_1,U_{1,k+1}^{n})\right.\nonumber\\
&&\quad\quad+H(Y_k|W_1,Y_1^{k-1})-H(Y_k|W_1,W_2,Y_1^{k-1},U_{2,k+1}^{n})-I(U_{2k};Y_1^{k-1}|W_1,W_2,U_{2,k+1}^{n})\nonumber\\
&&\quad\quad\left.-H(U_{1k})+H(U_{1k}|W_1,U_{1,k+1}^{n})-H(U_{2k})+H(U_{2k}|W_1,W_2,U_{2,k+1}^{n})\right]\nonumber\\
&=&\sum_{k=1}^{n}\left[H(Y_k|Y_1^{k-1})-H(Y_k|W_1,Y_1^{k-1},U_{1,k+1}^{n})\right.\nonumber\\
&&\quad\quad+H(Y_k|W_1,Y_1^{k-1})-H(Y_k|W_1,W_2,Y_1^{k-1},U_{2,k+1}^{n})\nonumber\\
&&\quad\quad\left.-H(U_{1k})+H(U_{1k}|W_1,U_{1,k+1}^{n},Y_1^{k-1})-H(U_{2k})+H(U_{2k}|W_1,W_2,U_{2,k+1}^{n},Y_1^{k-1})\right]\nonumber\\
&= &\sum_{k=1}^{n}\left[I(Y_k;W_1,U_{1,k+1}^{n}|Y_1^{k-1})+I(Y_k;W_2,U_{2,k+1}^{n}|W_1,Y_1^{k-1})\right.\nonumber\\
&&\quad\quad\left.-I(U_{1k};W_1,U_{1,k+1}^{n},Y_1^{k-1})-I(U_{2k};W_2,U_{2,k+1}^{n},Y_1^{k-1})\right]\nonumber\\
&=&\sum_{k=1}^{n}\left[H(W_1,U_{1,k+1}^{n}|Y_1^{k-1})-H(W_1,U_{1,k+1}^{n}|Y_1^{k-1},Y_k)\right.\nonumber\\
&&\quad\quad+H(W_2,U_{2,k+1}^{n}|W_1,Y_1^{k-1})-H(W_2,U_{2,k+1}^{n}|W_1,Y_1^{k-1},Y_k)\nonumber\\
&&\quad\quad-\left.I(U_{1k};W_1,U_{1,k+1}^{n},Y_1^{k-1})-I(U_{2k};W_2,U_{2,k+1}^{n},Y_1^{k-1})\right]\nonumber\\
&\stackrel{(a)}{=}&\sum_{k=1}^{n}\left[H(W_1,U_{1,k+1}^{n}|Y_1^{k-1})-H(W_1,U_{1,k+1}^{n}|Y_1^{k-1},Y_k)\right.\nonumber\\
&&\quad\quad+H(W_2,U_{2,k+1}^{n}|W_1,U_{1,k+1}^{n},Y_1^{k-1})-H(W_2,U_{2,k+1}^{n}|W_1,U_{1,k+1}^{n},Y_1^{k-1},Y_k)\nonumber\\
&&\quad\quad-\left.I(U_{1k};W_1,U_{1,k+1}^{n},Y_1^{k-1})-I(U_{2k};W_2,U_{2,k+1}^{n},Y_1^{k-1})\right]\nonumber\\
&=&\sum_{k=1}^{n}\left[H(W_1,U_{1,k+1}^{n},W_2,U_{2,k+1}^{n}|Y_1^{k-1})-H(W_1,U_{1,k+1}^{n},W_2,U_{2,k+1}^{n}|Y_1^{k-1},Y_k)\right.\nonumber\\
&&\quad\quad-\left.I(U_{1k};W_1,U_{1,k+1}^{n},Y_1^{k-1})-I(U_{2k};W_2,U_{2,k+1}^{n},Y_1^{k-1})\right]\nonumber\\
&=&\sum_{k=1}^{n}\left[I(W_1,U_{1,k+1}^{n},W_2,U_{2,k+1}^{n};Y_k|Y_1^{k-1})-I(U_{1k};W_1,U_{1,k+1}^{n},Y_1^{k-1})-I(U_{2k};W_2,U_{2,k+1}^{n},Y_1^{k-1})\right]\nonumber\\
&\leq &\sum_{k=1}^{n}\left[H(Y_k)-H(Y_k|W_1,U_{1,k+1}^{n},W_2,U_{2,k+1}^{n},Y_1^{k-1})-I(U_{1k};W_1,U_{1,k+1}^{n},Y_1^{k-1})\right.\nonumber\\
&&\qquad\qquad\left.-I(U_{2k};W_2,U_{2,k+1}^{n},Y_1^{k-1})\right]\nonumber\\
&= &\sum_{k=1}^{n}\left[I(W_1,U_{1,k+1}^{n},W_2,U_{2,k+1}^{n},Y_1^{k-1};Y_k)-I(U_{1k};W_1,U_{1,k+1}^{n},Y_1^{k-1})-I(U_{2k};W_2,U_{2,k+1}^{n},Y_1^{k-1})\right]\nonumber\\
&=&
\sum_{k=1}^{n}\left[I(L_{1k},L_{2k};Y_k)-I(U_{1k};L_{1k})-I(U_{2k};L_{2k})\right]\nonumber
\end{eqnarray}
where (a) holds since ($W_1,U_{1,k+1}^{n}$) is independent of
($W_2,U_{2,k+1}^{n}$) and $L_{1k}\triangleq
(W_1,Y_1^{k-1},U_{1,k+1}^{n})$ and $L_{2k}\triangleq
(W_2,Y_1^{k-1},U_{2,k+1}^{n})$ in the last equality. The above
implies
\[
R_1+R_2\leq
\frac{1}{n}\sum_{k=1}^{n}\left[I(L_{1k},L_{2k};Y_k)-I(U_{1k};L_{1k})-I(U_{2k};L_{2k})\right]+\epsilon_n.
\]
The rest of the proof proceeds the same way as the proof of Theorem
\ref{theo2}.\qed

\section{Upper Bounds on $|\mathcal{T}_i|$ for $\mathcal{R}^*_{in}(D_1,D_2)$ and $\mathcal{R}_{in}(D_1,D_2)$}\label{boundcar}

We only bound the cardinality of $\mathcal{T}_1$ and $\mathcal{T}_2$
for the region $\mathcal{R}^*_{in}(D_1,D_2)$. The bounds for
$|\mathcal{T}_1|$ and $|\mathcal{T}_2|$ for the region
$\mathcal{R}_{in}(D_1,D_2)$ can be derived in a similar manner. We
will need the following support lemma, which is based on
Carath\'{e}odory's theorem on the convex hull of a set  in
a finite-dimensional vector space. 

\begin{lemma}
{\rm (\cite[Support lemma, p.~311]{Csiszar3}) Let
$\mathcal{P}(\mathcal{X})$ be the set of distributions defined on a
finite set $\mathcal{X}$ (represented as the probability simplex in
$\mathbb{R}^{|\mathcal{X}|}$) and let $f_j$, $j=1,2,...,k$ be real-valued
continuous functions on $\mathcal{P}(\mathcal{X})$. For any probability measure
$\mu$ on the Borel $\sigma$-algebra of $\mathcal{P}(\mathcal{X})$,
there exist $k$ elements $P_1, P_2,..., P_k$ of
$\mathcal{P}(\mathcal{X})$ and $k$ non-negative reals
$\alpha_1,\alpha_2,...\alpha_k$ with $\sum_{i=1}^k \alpha_{i}=1$
such that for every $j=1,2,...,k$
$$
\int_{\mathcal{P}(\mathcal{X})}f_j(P)\mu(dP)=\sum_{i=1}^k
\alpha_if_{j}(P_{i}).
$$
}
\end{lemma}

Using this lemma, we will show that for any given $P_{X_1T_1|U_1}$
and $P_{X_2T_2|U_2}$, there exists a RV $\widehat{T}_1$ with
$|\widehat{\mathcal{T}}_1|\leq |\mathcal{U}_1||\mathcal{X}_1|+1$
only depending on $U_1$ and $X_1$ such that the following hold
\begin{eqnarray}
I(\widehat{T}_1;Y|T_2)-I(U_1;\widehat{T}_1)&=& I(T_1;Y|T_2)-I(U_1;T_1)\label{mutual1}\\
I(T_2;Y|\widehat{T}_1)-I(U_2;T_2)&=& I(T_2;Y|T_1)-I(U_2;T_2)\label{mutual2}\\
I(\widehat{T}_1,T_2;Y)-I(U_1;\widehat{T}_1)-I(U_2;T_2)&=&
I(T_1,T_2;Y)-I(U_1;T_1)-I(U_2;T_2),\label{mutual3}
\end{eqnarray}
and that the expectation of the distortion between $U_1$ and $X_1$
is preserved when $T_1$ is replaced by $\widehat{T}_1$. Note that
the upper bound on $|\widehat{\mathcal{T}}_1|$ does not depend on
$|\mathcal{T}_2|$.

We first rewrite
\begin{eqnarray}
I(T_1;Y|T_2)-I(U_1;T_1)&=&H(Y|T_2)-H(Y|T_1,T_2)-H(U_1)+H(U_1|T_1),
\nonumber\\
I(T_2;Y|T_1)-I(U_2;T_2)&=&H(Y|T_1)-H(Y|T_1,T_2)-I(U_2;T_2),\nonumber
\end{eqnarray}
and
$$
I(T_1,T_2;Y)-I(U_1;T_1)-I(U_2;T_2)=H(Y)-H(Y|T_1,T_2)-H(U_1)+H(U_1|T_1)-I(U_2;T_2).
$$
Recall that the joint distribution of $(U_1,U_2,T_2,T_2,X_1,X_2,Y)$
can be factorized as
$$
P_{U_1T_1U_2T_2X_1X_2Y}=Q_{U_1U_2}P_{T_1X_1|U_1}P_{T_2X_2|U_2}W_{Y|X_1X_2}.
$$
We note that there exists a Markov chain $(T_1,X_1)\rightarrow
U_1\rightarrow U_2\rightarrow (T_2,X_2)$. Writing
$$
P_{U_1T_1U_2T_2X_1X_2Y}=P_{T_1}P_{U_1X_1|T_1}P_{U_2|U_1}P_{T_2X_2|U_2}W_{Y|X_1X_2},
$$
and noting that $P_{U_2|U_1}$, $P_{T_2X_2|U_2}$ and $W_{Y|X_1X_2}$
are fixed, to apply the support lemma, we need $m-1$ functions to
preserve the joint distribution of $(U_1,X_1)$ (see (\ref{p1})
below), where $m\triangleq |\mathcal{U}_1||\mathcal{X}_1|$.
Specifically, we define the following real-valued continuous
functions of distribution $P_{U_1X_1|T_1}(\cdot,\cdot|t_1)$ on
$\mathcal{U}_1\times\mathcal{X}_1$ for fixed $t_1\in\mathcal{T}_1$,
$$
f_{u_1,x_1}(P_{U_1X_1|T_1}(\cdot,\cdot|t_1))\triangleq
P_{U_1X_1|T_1}(u_1,x_1|t_1)
$$
for all $(u_1,x_1)\in\mathcal{U}_1\times\mathcal{X}_1$ except one
pair $(u_1,x_1)$. Furthermore, we define real-valued continuous
functions
\begin{eqnarray}
f_{m}(P_{U_1X_1|T_1}(\cdot,\cdot|t_1)) &\triangleq&
-H_P(Y|T_1=t_1,T_2)+H_P(U_1|T_1=t_1),\nonumber\\
f_{m+1}(P_{U_1X_1|T_1}(\cdot,\cdot|t_1)) &\triangleq&
H_P(Y|T_1=t_1)-H_P(Y|T_1=t_1,T_2),\nonumber
\end{eqnarray}
where the entropies are taken under the joint distribution induced
by $P_{U_1X_1|T_1}(\cdot,\cdot|t_1)$.
According to the support lemma, there must exist a new RV
$\widehat{T}_1$ (jointly distributed with $(U_1,X_1)$) with alphabet
size $|\widehat{T}_1|=m+1=|\mathcal{U}_1||\mathcal{X}_1|+1$ such
that the expectation of $f_i$, $i=1,2,...,m+1$, with respect to
$P_{T_1}$ can be expressed in terms of the convex combination of
$m+1$ points; i.e.,
\begin{eqnarray}
P_{U_1X_1}(u_1,x_1)&=&\sum_{t_1\in\mathcal{T}_1}P_{T_1}(t_1)f_{u_1,x_1}(P_{U_1X_1|T_1}(\cdot,\cdot|t_1))\nonumber\\
&=&\sum_{\widehat{t}_1\in\widehat{\mathcal{T}}_1}P_{\widehat{T}_1}(\widehat{t}_1)f_{u_1,x_1}(P_{U_1X_1|\widehat{T}_1}(\cdot,\cdot|\widehat{t}_1)),\label{p1}
\end{eqnarray}
\begin{eqnarray}
-H(Y|T_1,T_2)+H(U_1|T_1)&=&\sum_{t_1\in\mathcal{T}_1}P_{T_1}(t_1)f_m(P_{U_1X_1|T_1}(\cdot,\cdot|t_1))\nonumber\\
&=&\sum_{\widehat{t}_1\in\widehat{\mathcal{T}}_1}P_{\widehat{T}_1}(\widehat{t}_1)f_m\left(P_{U_1X_1|\widehat{T}_1}(\cdot,\cdot|\widehat{t}_1)\right)\nonumber\\
&=&-H(Y|\widehat{T}_1,T_2)+H(U_1|\widehat{T}_1),\nonumber
\end{eqnarray}
\begin{eqnarray}
H(Y|T_1)-H(Y|T_1,T_2)&=&\sum_{t_1\in\mathcal{T}_1}P_{T_1}(t_1)f_{m+1}(P_{U_1X_1|T_1}(\cdot,\cdot|t_1))\nonumber\\
&=&\sum_{\widehat{t}_1\in\widehat{\mathcal{T}}_1}P_{\widehat{T}_1}(\widehat{t}_1)f_{m+1}(P_{U_1X_1|\widehat{T}_1}(\cdot,\cdot|\widehat{t}_1))\nonumber\\
&=&H(Y|\widehat{T}_1)-H(Y|\widehat{T}_1,T_2).\nonumber
\end{eqnarray}
This implies that (\ref{mutual1})--(\ref{mutual3}) hold. It should
be point out that this RV $\widehat{T}_1$ maintains the prescribed
distortion level, since $P_{U_1X_1}(u_1,x_1)$ is preserved.
Similarly, for any given $P_{X_1T_1|U_1}$ and $P_{X_2T_2|U_2}$, we
can show that there exists a RV $\widehat{T}_2$ with
$|\widehat{\mathcal{T}}_2|\leq |\mathcal{U}_2||\mathcal{X}_2|+1$
only depending on $U_2$ and $X_2$ such that
\begin{eqnarray}
I(T_1;Y|\widehat{T}_2)-I(U_1;T_1)&=& I(T_1;Y|T_2)-I(U_1;T_1)\label{mutual4}\\
I(\widehat{T}_2;Y|T_1)-I(U_2;\widehat{T}_2)&=& I(T_2;Y|T_1)-I(U_2;T_2)\label{mutual5}\\
I(T_1,\widehat{T}_2;Y)-I(U_1;T_1)-I(U_2;\widehat{T}_2)&=&
I(T_1,T_2;Y)-I(U_1;T_1)-I(U_2;T_2),\label{mutual6}
\end{eqnarray}
and the distortion constraint between $U_2$ and $X_2$ is preserved.
Thus we conclude that the cardinality of $\mathcal{T}_i$ can be
bounded by $|\mathcal{U}_i||\mathcal{X}_i|+1$, $i=1,2$.

Finally, we remark that the support lemma cannot be straightforwardly
used to bound 
the cardinality for $\mathcal{T}_1$ and $\mathcal{T}_2$ for the
region $\mathcal{R}_{out}(D_1,D_2)$ and
$\mathcal{R}^*_{out}(D_1,D_2)$. For example, to bound the
cardinality of $\mathcal{T}_1$ for $\mathcal{R}_{out}(D_1,D_2)$, we
need
$|\mathcal{U}_1||\mathcal{U}_2||\mathcal{X}_1||\mathcal{X}_2||\mathcal{T}_2|-1$
real-valued continuous functions to preserve the joint distribution
of $(U_1,U_2,T_2,X_1,X_2)$. Therefore, we may need
$|\mathcal{U}_1||\mathcal{U}_2||\mathcal{X}_1||\mathcal{X}_2||\mathcal{T}_2|+1$
letters and this upper bound depends on $|\mathcal{T}_2|$. \qed


\end{document}